\documentclass[10pt, conference, compsocconf]{IEEEtran}

\newcommand{\nc}{\textit{Neural Cache}}

\usepackage{fancyhdr}
\usepackage{microtype}

\usepackage{cite}

\usepackage[bookmarks=true,breaklinks=true,letterpaper=true,colorlinks,linkcolor=black,citecolor=blue,urlcolor=black]{hyperref}

\usepackage{textcomp}
\usepackage{color}
\usepackage[labelfont=bf, textfont=bf]{caption}   

\usepackage[]{subfig}
\usepackage{array}

\usepackage[moderate]{savetrees}

\fancypagestyle{firstpage}{
  \fancyhf{}
\setlength{\headheight}{50pt}

  \fancyhead[C]{\normalsize{To appear in the 45th ACM/IEEE International Symposium on Computer Architecture (ISCA 2018)}}
  \pagenumbering{arabic}
}  

\usepackage[pdftex]{graphicx}
\graphicspath{{./figs/}}

\usepackage{mdwmath}
\usepackage{mdwtab}

\begin{document}

%
\title{Neural Cache: Bit-Serial In-Cache Acceleration of Deep Neural Networks}

\author{
Charles Eckert, Xiaowei Wang, Jingcheng Wang, Arun Subramaniyan,
\\Ravi Iyer$^\dagger$, Dennis Sylvester, David Blaauw, and Reetuparna Das\\
University of Michigan \qquad $^\dagger$Intel Corporation\\
\{eckertch, xiaoweiw, jiwang, arunsub, dmcs, blaauw, reetudas\}@umich.edu, ravishankar.iyer@intel.com}



\maketitle

\thispagestyle{firstpage}






\begin{abstract}
	This paper presents the \nc{} architecture, which re-purposes cache structures to transform them into massively parallel compute units capable of running inferences for Deep Neural Networks. Techniques to do in-situ arithmetic in SRAM arrays, create efficient data mapping and reducing data movement are proposed. The \nc{} architecture is capable of fully executing convolutional, fully connected, and pooling layers in-cache. The proposed architecture also supports quantization in-cache. 
    
    Our experimental results show that the proposed architecture can improve inference latency by 18.3$\times$ over state-of-art multi-core CPU (Xeon E5), 7.7$\times$ over server class GPU (Titan Xp), for Inception v3 model. \nc{} improves inference throughput by 12.4$\times$ over CPU  (2.2$\times$ over GPU), while reducing power consumption by 50\% over CPU (53\% over GPU).

 \end{abstract}

\section{Introduction}

In the last two decades,  the number of processor cores per chip has steadily increased while memory latency has remained relatively constant. This has lead to the so-called memory wall~\cite{mwall} where 
memory bandwidth and memory energy have come to dominate computation bandwidth and energy. With the advent of data-intensive system, this problem is further exacerbated and as a result, today a large fraction of energy is spent in moving data back-and-forth between memory and compute units. At the same time, neural computing and other data intensive computing applications have emerged as increasingly popular applications domains, exposing much higher levels of data parallelism. In this paper, we exploit both these synergistic trends by \emph{opportunistically} leveraging the huge caches present in modern processors to perform massively parallel processing for neural computing.   

Traditionally, researchers have attempted to address the memory wall by building a deep memory hierarchy. Another solution is to move compute closer to memory, which is often referred to as processing-in-memory (PIM). Past PIM~\cite{Gokhale95,Kogge94,Patterson97} solutions tried to move computing logic {\em near} DRAM by integrating DRAM with a logic die using 3D stacking~\cite{Hmc14, Ahn15, neurocube-isca16}. This helps reduce latency and increase bandwidth, however, the functionality and design of DRAM itself remains unchanged. Also, this approach adds substantial cost to the overall system as each DRAM die needs to be augmented with a separate logic die. Integrating computation on the DRAM die itself is difficult since the DRAM process is not optimized for logic computation.

In this paper, we instead completely eliminate the line that distinguishes memory from compute units. Similar to the human brain, which does not separate these two functionalities distinctly, we perform computation directly on the bit lines of the memory itself, keeping data in-place. This eliminates data movement and hence significantly improves energy efficiency and performance. Furthermore, we take advantage of the fact that over 70\% of silicon in today's processor dies simply stores and provides data retrieval; harnessing this area by re-purposing it to perform computation can lead to massively parallel processing. 

The proposed approach builds on an earlier silicon test chip implementation~\cite{jeloka201628} and architectural prototype~\cite{cc-hpca17} that shows how simple logic operations (AND/NOR) can be performed directly on the bit lines in a 
standard SRAM array. This is performed by enabling SRAM rows simultaneously while leaving the operands in-place in memory. 
This paper presents the \nc{} architecture which leverages these simple logic operations to perform \textit{arithmetic computation} (add, multiply, and reduction) \textit{directly in the SRAM array} by storing the data in transposed form and performing bit-serial computation while incurring only an estimated 7.5\% area overhead (translates to less than 2\% area overhead for the processor die). Each column in an array performs a separate calculation and the thousands of memory arrays in the cache can operate concurrently. 

The end result is that cache arrays morph into massive vector compute units (up to 1,146,880 bit-serial ALU slots in a Xeon E5 cache) that are one to two orders of magnitude larger than modern graphics processor's (GPU's) aggregate vector width. By avoiding data movement in and out of memory arrays, we naturally save vast amounts of energy that is typically spent in shuffling data between compute units and on-chip memory units in modern processors. 

 \nc{} leverages opportunistic in-cache computing resources for accelerating Deep Neural Networks (DNNs). There are two key challenges to harness a cache's computing resources. First, all the operands participating in an in-situ operation must share bit-lines and be mapped to the same memory array. Second, intrinsic data parallel operations in DNNs have to be exposed to the underlying parallel hardware and cache geometry. We propose a data layout and execution model that solves these challenges, and harnesses the full potential of in-cache compute capabilities. Further, we find that thousands of in-cache compute units can be utilized by replicating data and improving data reuse. Techniques for low-cost replication, reducing data movement overhead, and improving data reuse are discussed.

In summary, this paper offers the following contributions:

\begin{itemize}
\item
This is the first work to demonstrate in-situ arithmetic compute capabilities for caches. We present a compute SRAM array design which is capable of performing additions and multiplications. A critical challenge for enabling complex operations in cache is facilitating interaction between bit lines.  We propose a novel bit-serial implementation with
transposed data layout to address the above challenge. We designed an 8T SRAM-based hardware transpose unit for dynamically transposing data on the fly.

\item Compute capabilities transform caches to massively data-parallel co-processors at negligible area cost. For example, we can re-purpose the 35 MB Last Level Cache (LLC) in the server class Intel Xeon E5 processor to support 1,146,880 bit-serial ALU slots. Furthermore, in-situ computation in caches naturally saves the on-chip data movement overheads.

\item
We present the \nc{} architecture which re-purposes last-level cache to accelerate DNN inferences. A key challenge is exposing the parallelism in DNN computation to the underlying cache geometry. We propose a data layout that solves these challenges, and harnesses the full potential of in-cache compute capabilities. Further, techniques which reduce data movement overheads are discussed.

\item 
The \nc{} architecture is capable of fully executing convolutional, fully connected, and pooling layers in-cache. The proposed architecture also supports quantization and normalization in-cache. Our experimental results show that the proposed architecture can improve inference latency by 18.3$\times$ over state-of-the-art multi-core CPU (Xeon E5), 7.7$\times$ over server class GPU (Titan Xp), for the Inception v3 model. \nc{} improves inference throughput by 12.4$\times$ over CPU  (2.2$\times$ over GPU), while reducing power consumption by 59\% over CPU (61\% over GPU).

\end{itemize}
\section{Background}
\subsection{Deep Neural Networks}
Deep Neural Networks (DNN) have emerged as popular machine learning tools. Of particular fame are Convolutional Neural Networks (CNN) which have been used for various visual recognition tasks ranging from object recognition and detection
to scene understanding. While \nc{} can accelerate the broader class of DNNs, this paper focuses on CNNs. 

\begin{figure}[h]
  
	\minipage{0.5\textwidth}
\centering
  	\includegraphics[scale=0.18]{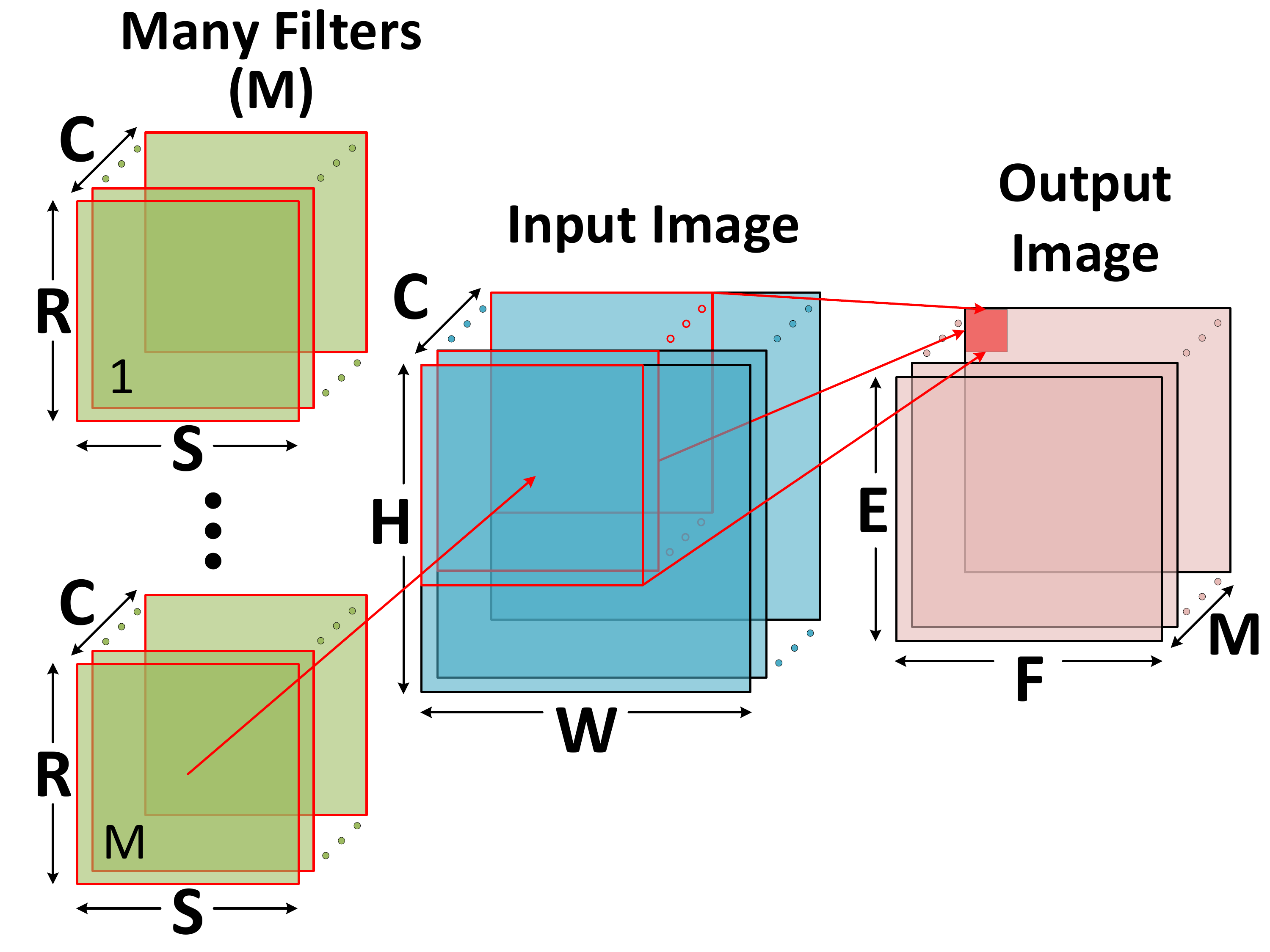}
	\endminipage
  \caption{Computation of a convolution layer.} 
	\label{fig:conv}
    \vspace{2mm}
\end{figure}

CNNs consist of multiple convolution layers with additional pooling, normalization, and fully connected layers mixed in. Since convolutional layers are known to account for over 90\% of inference operations~\cite{cong2014minimizing, chen2016eyeriss}, we discuss them in this section. A convolutional layer performs a convolution on the filter and the input image. Filters have four dimensions, a height (R), width (S), channels (C), and batches of 3D filters (M). Inputs have three dimensions with a height (H), width (W), and channel (C). The filter is overlaid on the inputs and each  pixel of the input is multiplied by the corresponding filter pixel and repeated across the M dimension. The results across the $R\times S$, i.e. the height and width, are accumulated together. Further, the channels are also reduced into a single element. Thus each window gives us an output of $1\times 1\times M$. The window is slid using a stride (U), increasing the stride will decrease the number of output elements computed. Eventually the sliding window will produce an output image with dimensions based on the height and width of the input, and the stride. The output's channel size is equivalent to the M dimension of the filter. The output image, after an activation function that varries across networks and layers, is fed as the input to the next layer.  
\begin{figure}[t]
          \subfloat[][]{
		\includegraphics[width=0.22\textwidth]{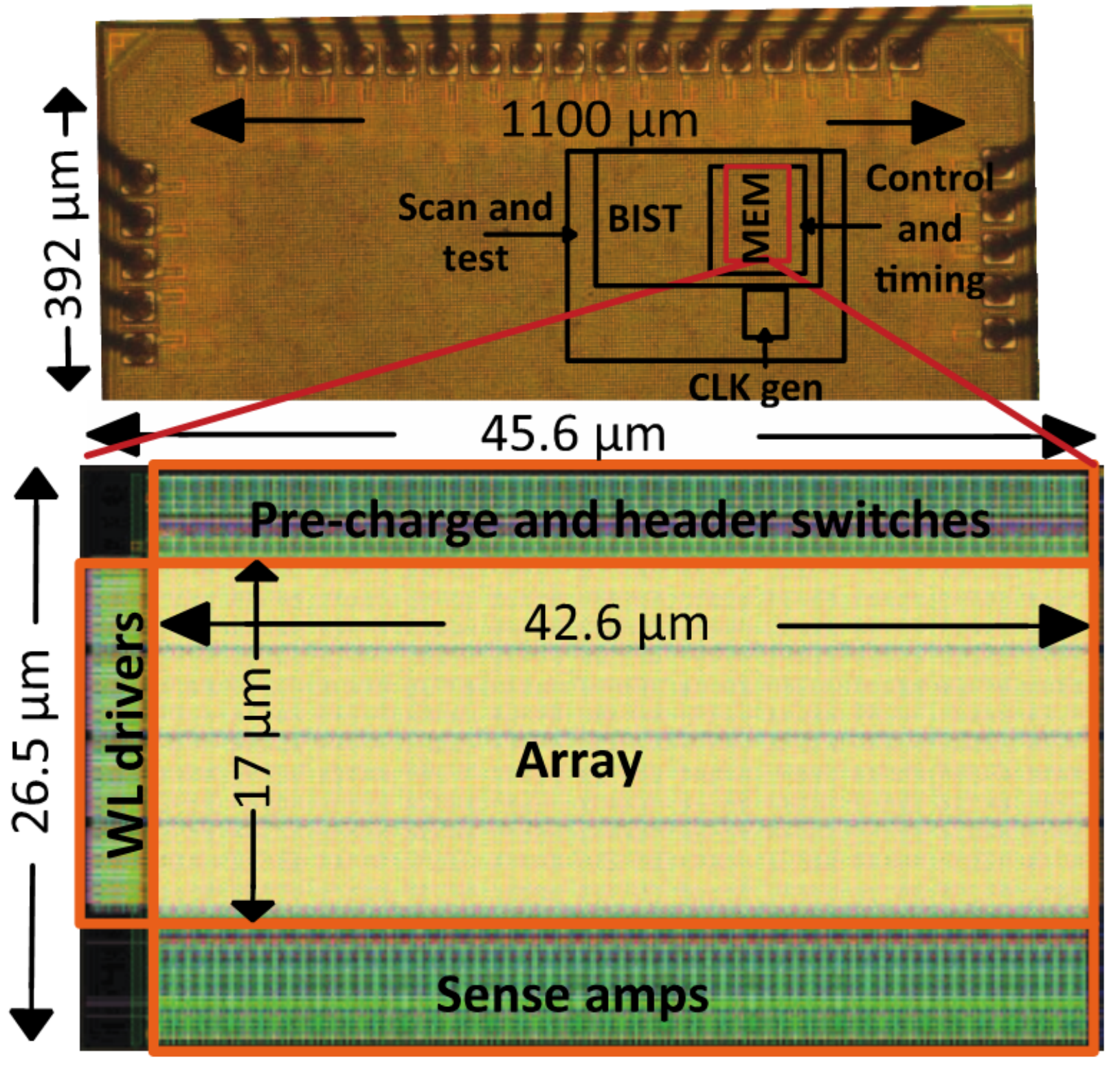}
     	\label{fig:chip}
     }
     \subfloat[][]{
        \includegraphics[width=0.24\textwidth]{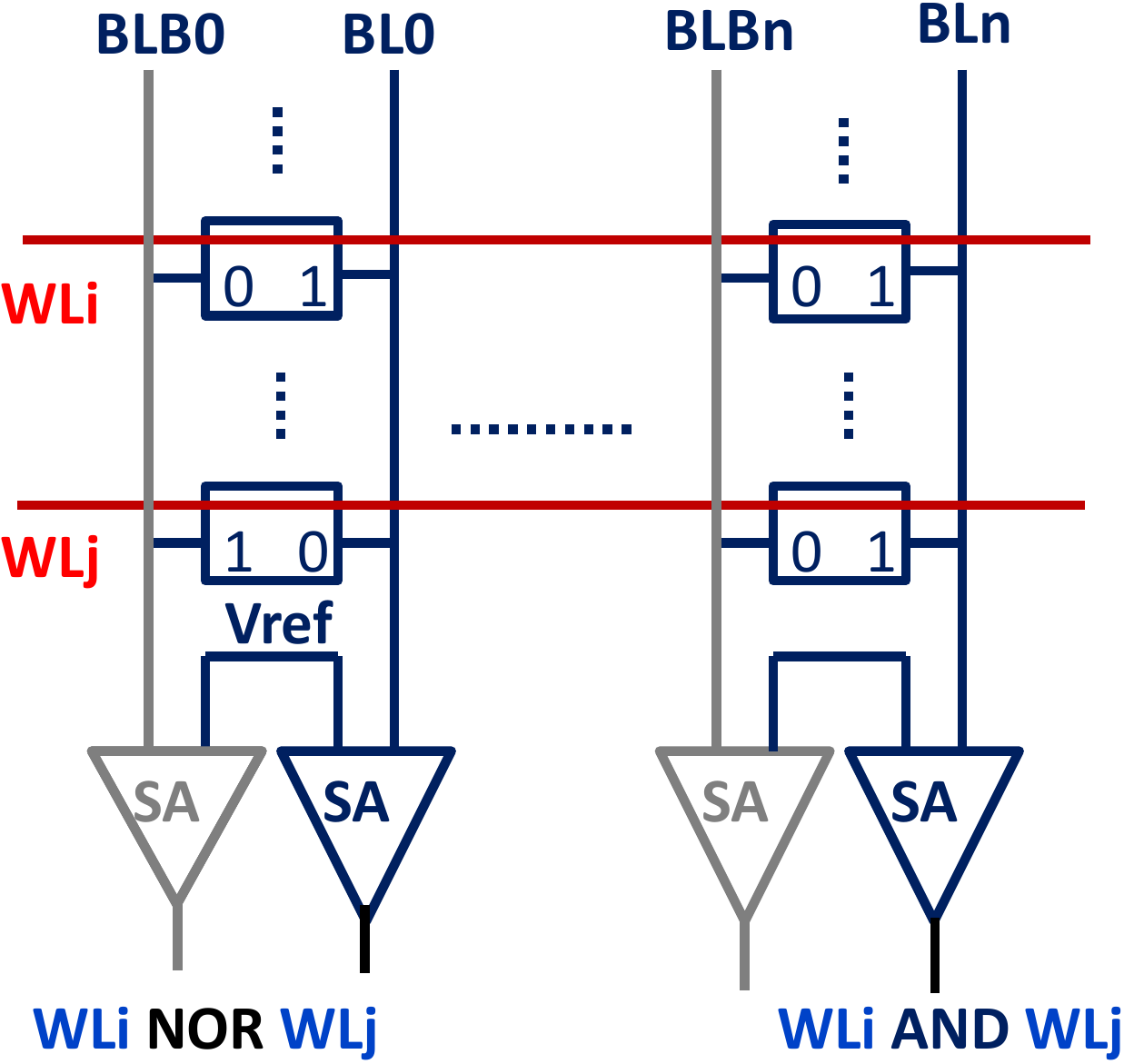}
        \label{fig:sramback}
     }

     \caption{(a) Prototype test chip~\cite{jeloka201628} (b) {SRAM circuit for in-place operations. Two rows (WL$_{i}$ and WL$_{j}$) are simultaneously activated. An \emph{AND} operation is performed by sensing bit-line (BL). A \emph{NOR} operation is performed by sensing bit-line complement (BLB).} }
     \vspace{-4mm}
\end{figure}

The convolutions performed can be broken down to many different dimensions of matrix-vector multiplications.
\nc{} breaks down a convolution to vector-vector dot product (vector dimension $R\times S$), followed by reduction across channels (C). The different filter batches (M) are computed in parallel using the above step. A new input vector is used for each stride. In this paper we analyze the state-of-art Inception v3 model which has 94 convolutional sub-layers. \nc{} is utilized to accelerate not only convolutional layers, but also pooling and fully connected layers.

\subsection{Bit-Line Computing}
\nc{} builds on earlier work on SRAM bit line circuit technology~\cite{jeloka201628,Jeloka15,Kang15} shown in Figure~\ref{fig:sramback}. To compute in-place, two word-lines are activated simultaneously. Computation ({\tt and} and {\tt nor}) on the data stored in the activated word-lines is performed in the analog domain by sensing the shared bit-lines. Compute cache~\cite{cc-hpca17}  uses this basic circuit framework along with  extensions to support additional operations: {\tt copy}, bulk zeroing, {\tt xor}, equality comparison, and search. 

Data corruption due to multi-row access is prevented by lowering the word-line voltage to bias against the write of the SRAM array. Measurements across 20 fabricated $28\; nm$ test chips  (Figure~\ref{fig:chip}) demonstrate that data corruption does not occur even when 64 word-lines are simultaneously activated during such an in-place computation. Compute cache however only needs two. Monte~Carlo simulations also show a stability of more than six sigma robustness, which is considered industry standard for robustness against process variations. The robustness comes at the the cost of increase in delay during compute operations. But, they have no effect on conventional array read/write accesses. The increased delay is more than compensated by massive parallelism exploited by \nc{}.

\begin{figure*}[t]
  
	\minipage{0.95\textwidth}
\centering
  	\includegraphics[scale=0.5]{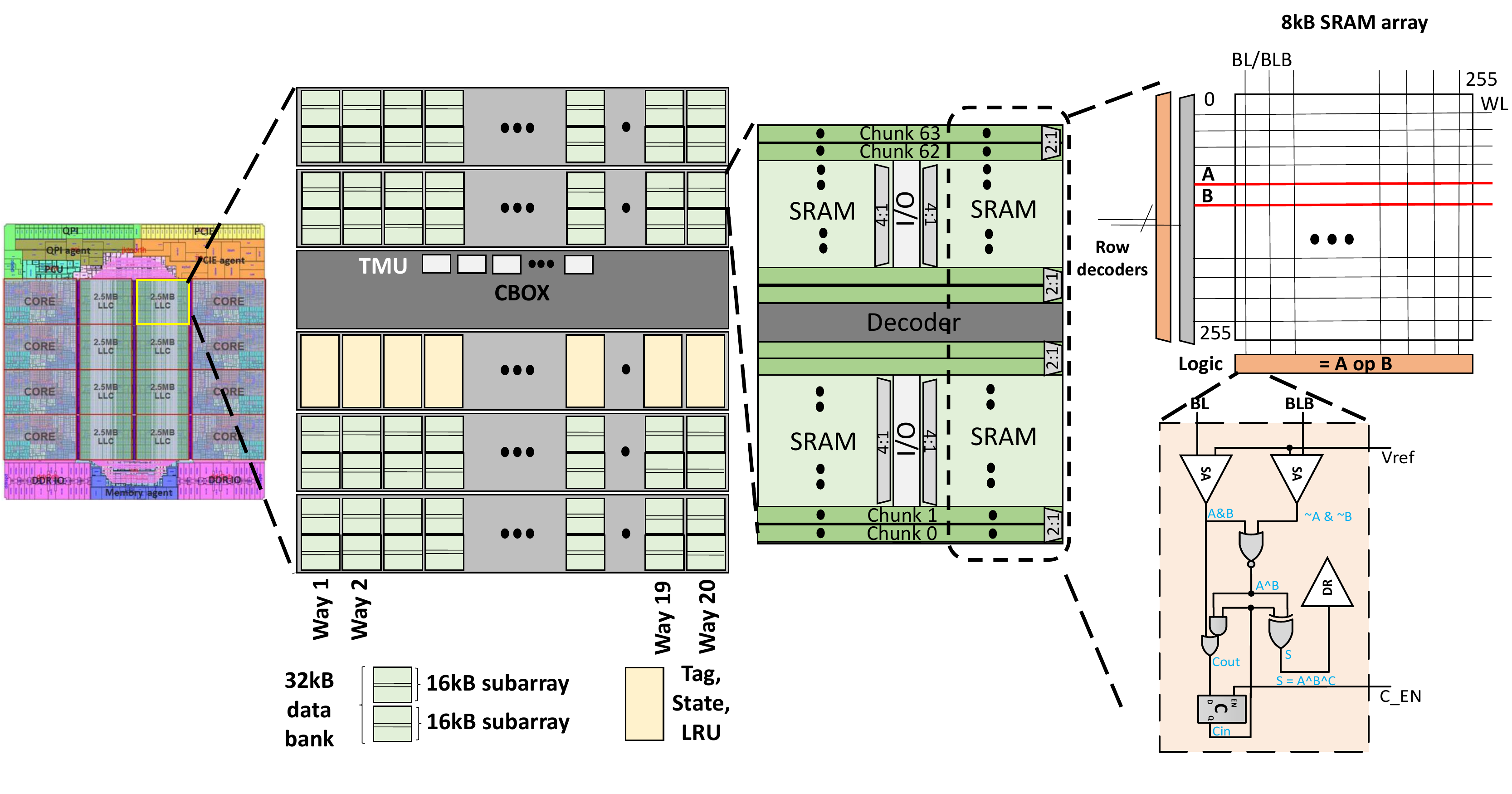}
	\endminipage
  \caption{Neural Cache overview. (a) Multi-core processor with 8-24 LLC slices. (b) Cache geometry of a single 2.5MB LLC cache slice
    with 80 32KB banks. Each 32KB bank has two 16KB
    sub-arrays.  (c) One 16KB sub-array composed of two 8KB SRAM
    arrays. (d) One 8KB SRAM array re-purposed to store data and do
    bit line computation on operand stored in rows (A and B). (e)
    Peripherals of the SRAM array to support computation.}
	\label{fig:cc}
    \vspace{4mm}
\end{figure*}
\subsection{Cache Geometry}

We provide a brief overview of a cache's geometry in a modern processor.
Figure~\ref{fig:cc} illustrates a multi-core processor modeled
loosely after Intel's Xeon processors~\cite{xeonllc45nm, xeonllc22nm}. Shared Last Level Cache (LLC) is distributed
into many slices (14 for Xeon E5 we modeled), which are accessible to the cores through a shared ring
interconnect (not shown in figure). Figure~\ref{fig:cc} (b) shows a slice of
 LLC cache. The slice has 80 32KB banks organized into 20 ways.  Each
 bank is connected by two 16KB  sub-arrays. Figure~\ref{fig:cc} (c)
 shows the internal organization of one 16KB sub-array, composed of 8KB
 SRAM arrays. Figure~\ref{fig:cc} (d) shows one 8KB SRAM array.
 A SRAM array is organized into multiple rows of
 data-storing bit-cells. Bit-cells in the
same row share one word line, whereas  bit-cells in the same column share one pair of
bit lines. 

Our proposal is to perform in-situ vector arithmetic operations within the
SRAM arrays (Figure~\ref{fig:cc} (d)). The resulting architecture can have
massive parallelism by repurposing thousands of SRAM arrays (4480 arrays in Xeon E5) into
vector computational units. 

We observe that LLC access latency is dominated by wire delays inside a cache slice, accessing upper-level cache control structures, and network-on-chip. Thus, while a typical LLC access can take $\sim$30 cycles, an SRAM array access is only 1 cycle (at 4 GHz clock~\cite{xeonllc45nm}).  Fortunately, in-situ architectures such as \nc{} require only SRAM array accesses and do not incur the overheads of a traditional cache access. Thus, vast amounts of energy and time
spent on wires and higher-levels of memory hierarchy can be saved.

\section{Neural Cache Arithmetic}
Compute cache~\cite{cc-hpca17} supported several simple operations (logical and copy). 
These operations are bit-parallel and do not require interaction between bit lines. \nc{} requires
support for more \emph{complex operations (addition,
multiplication, reduction)}.  The critical challenge in
supporting these complex computing primitives is facilitating interaction between bit lines. Consider
supporting an addition operation which requires carry propagation
between bit lines. We propose \textbf{bit-serial implementation} with
\textbf{transposed data layout} to address the above challenge.
\begin{figure*}[htb]
  \centering
	\minipage{0.66\textwidth}
  	\includegraphics[scale=0.45]{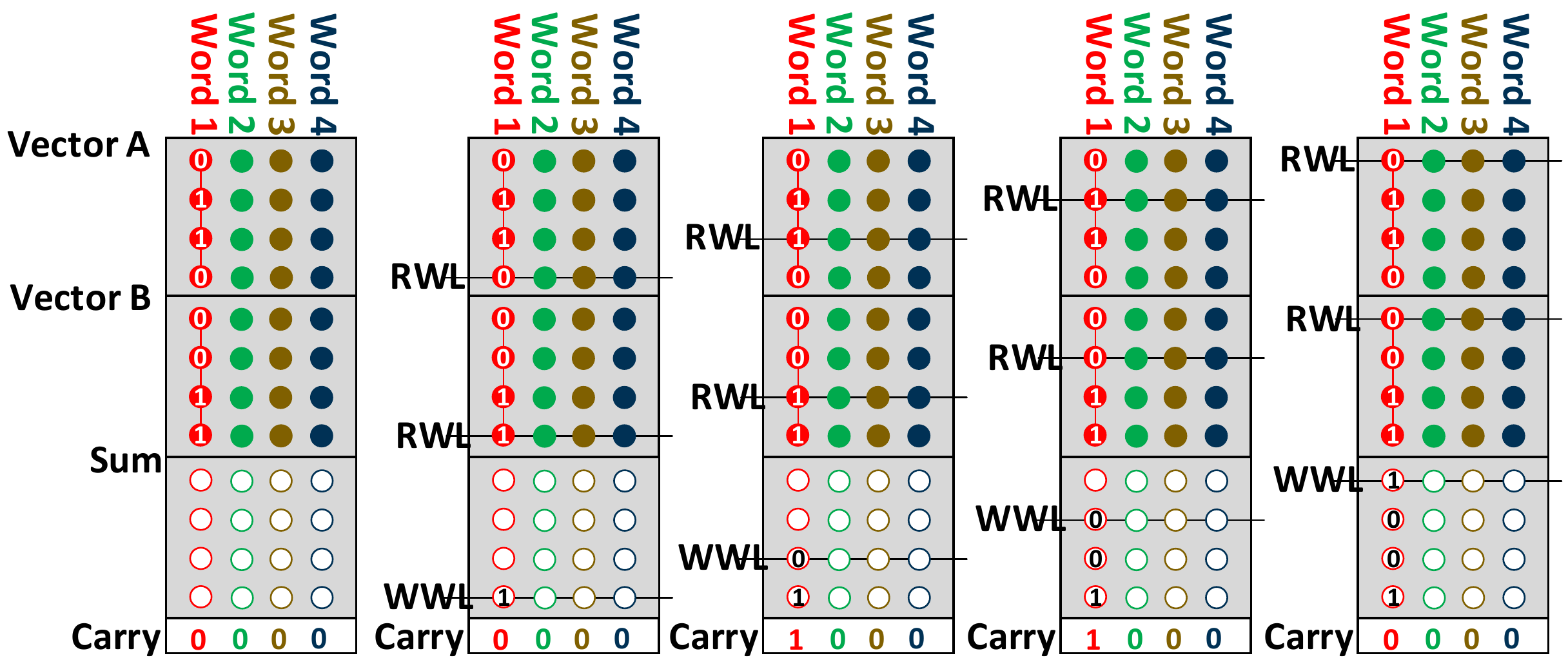}
    \caption{Addition operation} 
    \label{fig:add}
	\endminipage
	\minipage{0.33\textwidth}
  	\includegraphics[scale=0.35]{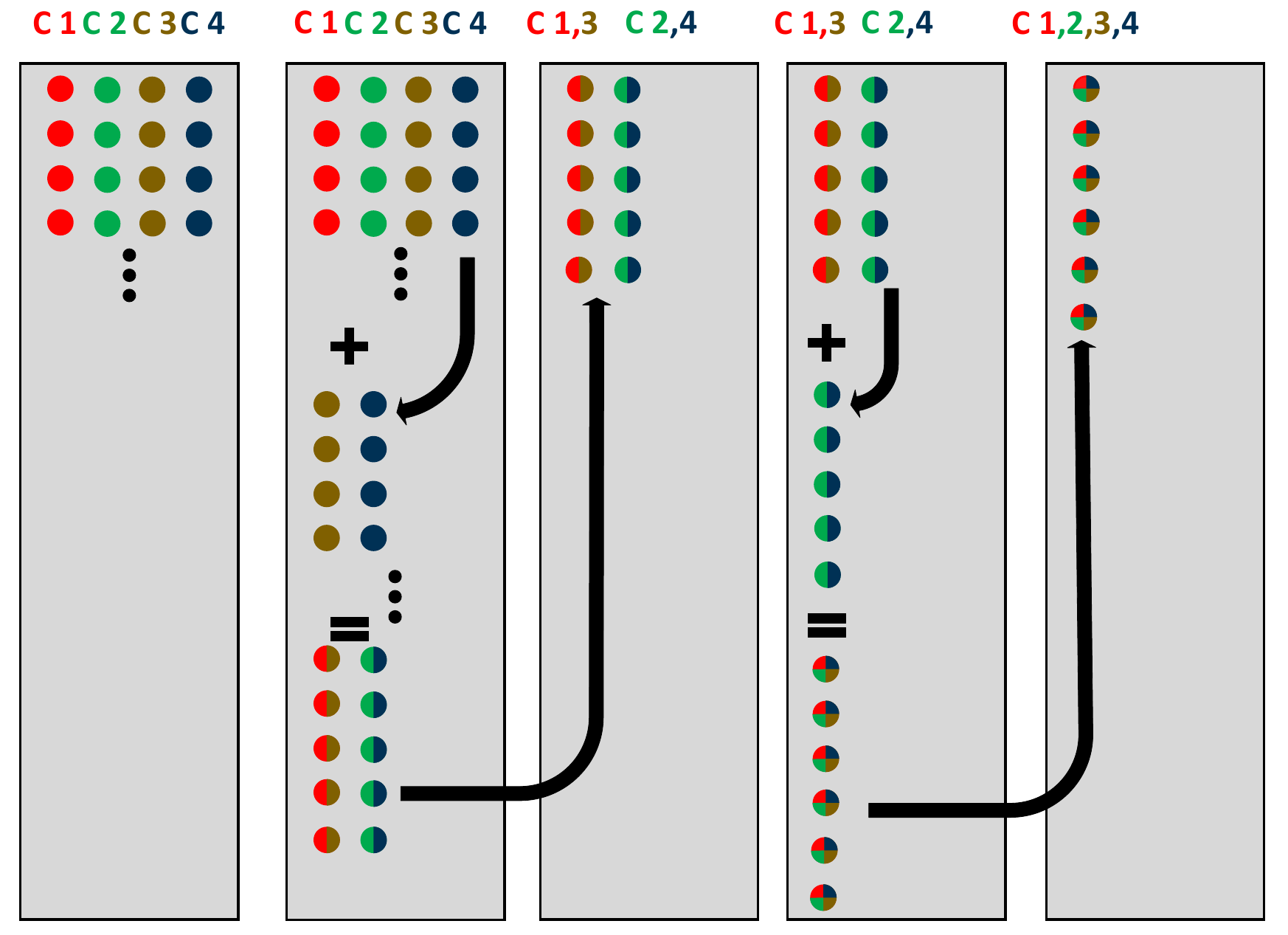}
    \caption{Reduction Operation} 
    \label{fig:reduce}
	\endminipage
	\vspace{2mm}
	\label{fig:ops}
\end{figure*}

\begin{figure*}[htb]
  \centering
	\minipage{0.5\textwidth}
  	\includegraphics[scale=0.45]{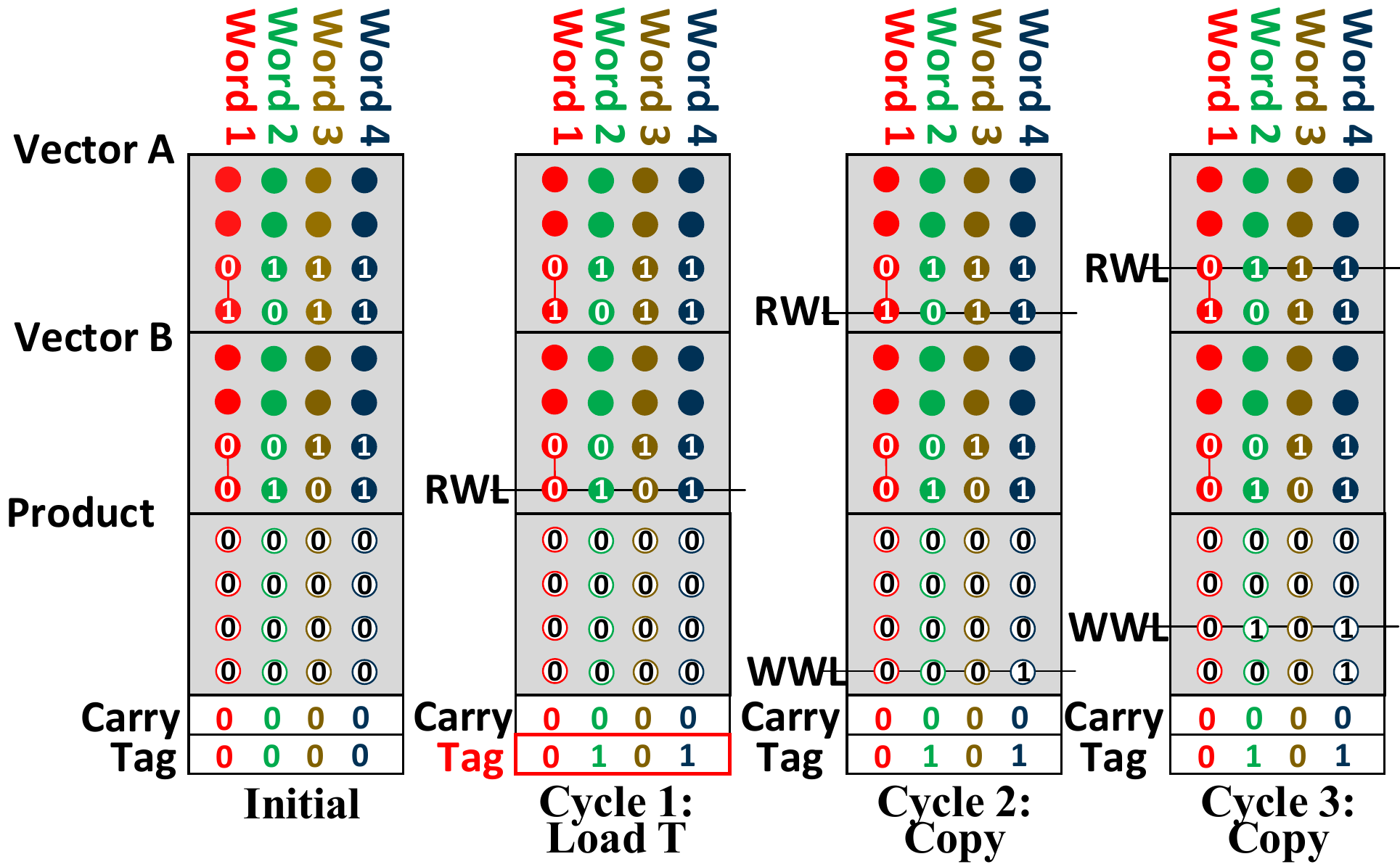}
	\endminipage
	\minipage{0.5\textwidth}
  	\includegraphics[scale=0.45]{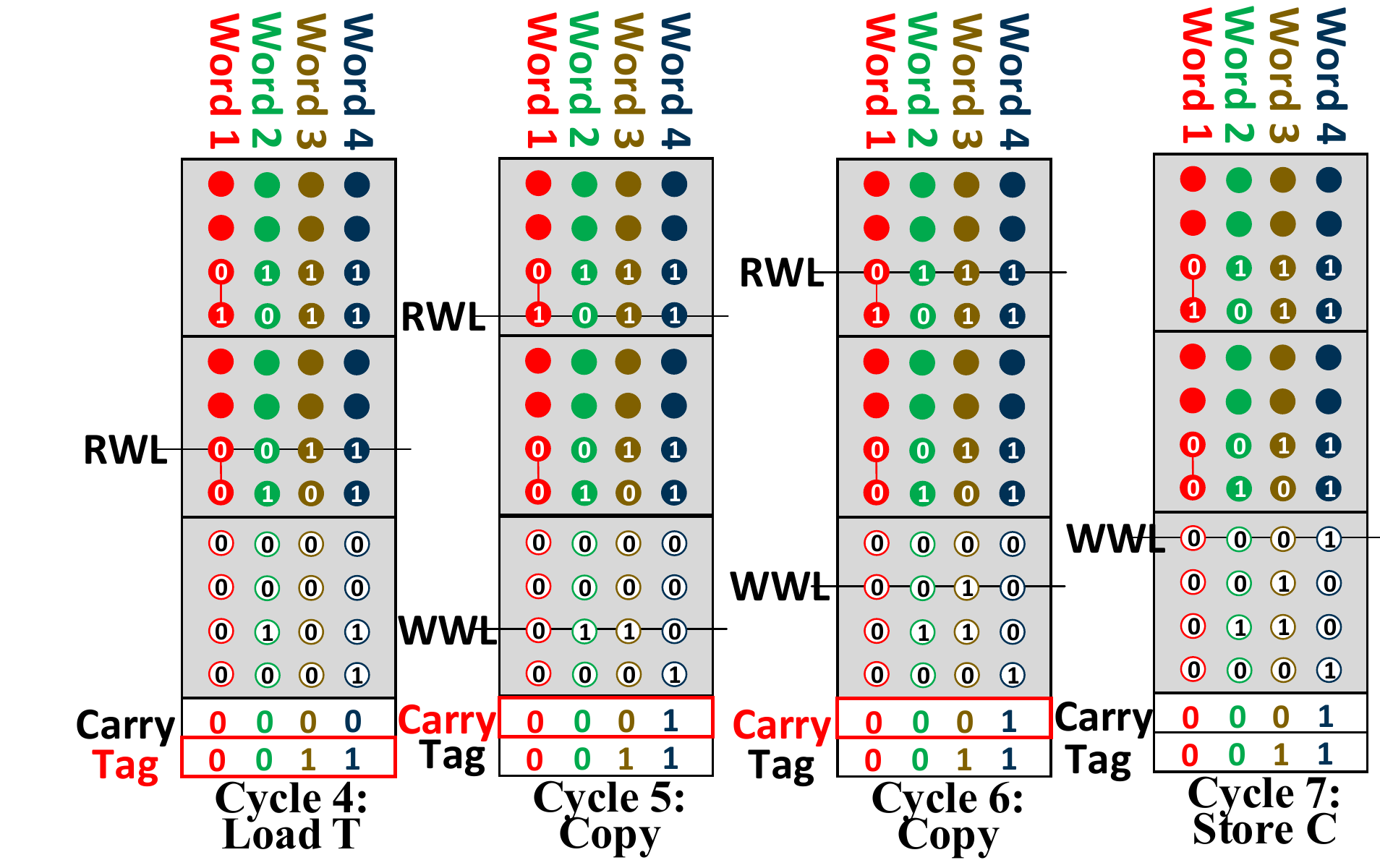}
	\endminipage
  \caption{Multiplication Operation, each step is shown with four different sets of operands} 
	\label{fig:mul}
      \vspace{2mm}
\end{figure*}
\subsection{Bit-Serial Arithmetic}
Bit-serial computing architectures have been widely used for digital
signal processing~\cite{bitserial1,bitserial2} because of their ability to provide
massive bit-level parallelism at low area costs. The key idea is to
process one bit of multiple data elements every cycle. This model is particularly
useful in scenarios where the same operation is applied to the same
bit of multiple data elements. Consider the following example to compute
the element-wise sum of two arrays with 512 32-bit elements. A conventional processor would process these arrays
element-by-element taking 512 steps to complete the operation. A
bit-serial processor, on the other hand, would complete the operation
in 32 steps as it processes the arrays \emph{bit-slice by bit-slice instead
of element-by-element}. Note that a bit-slice is composed of bits from
the same bit position, but corresponding to different elements of the
array. Since the number of elements in arrays is typically much
greater than the bit-precision for each element stored in them,
bit-serial computing architectures can provide much higher throughput
than bit-parallel arithmetic. Note also that bit-serial operation allows for 
flexible operand bit-width, which can be advantageous in DNNs 
where the required bit width can vary from layer to layer. 

Note that although bit-serial
computation is expected to have higher latency per operation,  it is expected to have significantly larger throughput, which
compensates for higher operation latency. For example, the 8KB SRAM
array is composed of 256 word lines and 256
bit lines and can operate at a maximum frequency of 4 GHz for accessing data~\cite{xeonllc45nm,xeonllc22nm}. Up to 256
elements can be processed in parallel in a single array. A 2.5 MB LLC slice has 320 8KB arrays as shown in Figure~\ref{fig:cc}. Haswell server processor's 35 MB LLC
 can accommodate 4480 such 8KB arrays. Thus up to 1,146,880
elements can be processed in parallel, while operating at frequency of 2.5 GHz when computing. By repurposing memory arrays, we gain the above throughput for
near-zero cost. Our circuit analysis estimates
an area overhead of additional bit line peripheral logic to be 7.5\% for each 8KB array. This translates to less than 2\% area overhead for the processor die.

\subsection{Addition}
\label{sec:add}
In conventional architectures, arrays are generated, stored, accessed,
and processed element-by-element in the vertical direction along the
bit lines. We refer to this data layout as the bit-parallel or regular data
layout. Bit-serial computing in SRAM arrays can be realized by storing data
elements in a transpose data layout. Transposing ensures that all
  bits of a data element are mapped to the same bit line, thereby obviating
  the necessity for communication between bit lines. Section~\ref{sec:transpose} discusses techniques to store data in a transpose layout. Figure~\ref{fig:add} shows an example $12 \times 4$ SRAM array with
transpose layout. The array stores two vectors A and B, each with four 4-bit elements. 
Four word lines are necessary to store all bit-slices of 4-bit elements.  
  
We use the addition of two vectors of 4-bit numbers to explain how addition works in the SRAM. The 2 words that are going to be added together have to be put in the same bit line. The vectors A and B should be aligned in the array like Figure~\ref{fig:add}. Vector A occupies the first 4 rows of the SRAM array and vector B the next 4 rows.  Another 4 empty rows of storage are reserved for the results. There is a row of latches inside the column peripheral for the carry storage. The addition algorithm is carried out bit-by-bit starting from the least significant bit (LSB) of the two words. There are two phases in a single operation cycle. In the first half of the cycle, two read word lines (RWL) are activated to simultaneously sense and wire-and the value in cells on the same bit line. To prevent the value in the bit cell from being disturbed by the sensing phase, the RWL voltage should be lower than the normal VDD. The sense amps and logic gates in the column peripheral (Section~\ref{sec:circuit}) use the 2 bit cells as operands and carry latch as carry-in to generate sum and carry-out. In the second half of the cycle, a write word line (WWL) is activated to store back the sum bit. The carry-out bit overwrites the data in the carry latch and becomes the carry-in of the next cycle. As demonstrated in Figure~\ref{fig:add}, in cycles 2, 3, and 4, we repeat the first cycle to add the second, third, and fourth bit respectively. Addition takes n + 1, to complete with the additional cycle to write a carry at the end.

\subsection{Multiplication}
\label{sec:mul}
We demonstrate how bit-serial multiplication is achieved based on addition and predication using the example of a 2-bit multiplication. In addition to the carry latch, an extra row of storage, the tag latch, is added to bottom of the array. The tag bit is used as an enable signal to the bit line driver. When the tag is one, the addition result sum will be written back to the array. If the tag is zero, the data in the array will remain. Two vectors of 2-bit numbers, A and B, are stored in the transposed fashion and aligned as shown in Figure~\ref{fig:mul}. Another 4 empty rows are reserved for the product and initialized to zero. Suppose A is a vector of multiplicands and B is a vector of multipliers. First, we load the LSB of the multiplier to the tag latch. If the tag equals one, the multiplicand in that bit line will be copied to product in the next two cycles, as if it is added to the partial product. Next, we load the second bit of the multiplier to the tag. If tag equals 1, the multiplicand in that bit line will be added to the second and third bit of the product in the next two cycles, as if a shifted version of A is added to the partial product. Finally, the data in the carry latch is stored to the most significant bit of the product. Including the initialization steps, it takes $n^2 + 5n -2$ cycles to finish an $n$-bit multiplication. Division can be supported using a similar algorithm and takes $1.5n^2 + 5.5n$ cycles. 


\subsection{Reduction}
\label{sec:red}
Reduction is a common operation for DNNs. Reducing the elements stored on different bit lines to one sum can be performed with a series of word line moves and additions. Figure~\ref{fig:reduce} shows an example that reduces 4 words, $C1$, $C2$, $C3$, and $C4$. First words $C3$ and $C4$ are moved below $C1$ and $C2$ to different word lines. This is followed by addition. Another set of move and addition reduces the four elements to one word. Each reduction step increases the number of word lines to move as we increase the bits for the partial sum. The number of reduction steps needed is $\log_{2}$ of the words to be reduced. In column multiplexed SRAM arrays, moves between word lines can be sped up using sense-amp cycling techniques~\cite{ca-micro17}. 

When the elements to be reduced do not fit in the same SRAM array, reductions must be performed across arrays which can be accomplished by inter-array moves. In DNNs, reduction is typically performed across channels. In the model we examined, our optimized data mapping is able to fit all channels in the space of two arrays which sharing sense amps. We employ a technique called packing that allows us to reduce the number of channels in large layers (Section~\ref{sec:dm}). 

\begin{figure}[h]
	\minipage{0.5\textwidth}
\centering
  	\includegraphics[scale=0.75]{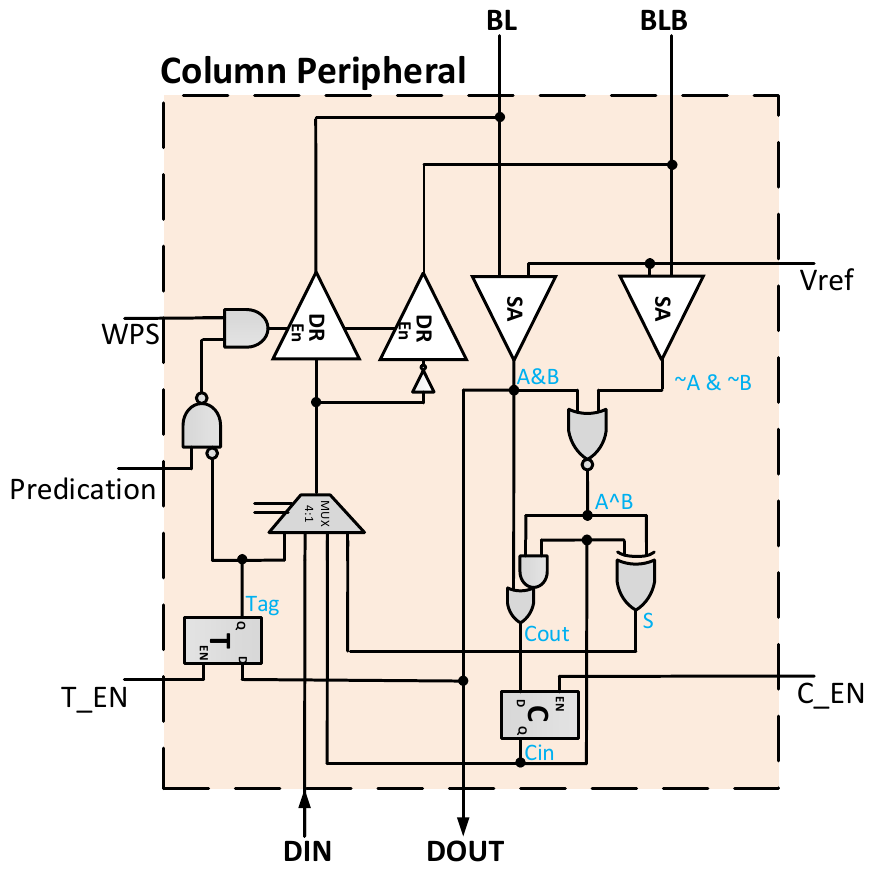}
	\endminipage
  \caption{Bit line peripheral design} 
	\label{fig:circuit}
\end{figure}
\subsection{SRAM Array Peripherals}
\label{sec:circuit}
The bit-line peripherals are shown in Figure~\ref{fig:circuit}. Two single-ended sense amps sense the wire-and result from two cells, A and B, in the same bitline. The sense amp in BL gives result of $A\; \& \; B$, while the sense amp in BLB gives result of 
$A'\; \& \;B'$. The sense amps can use reconfigurable sense amplifier design~\cite{Jeloka15}, which can combine into a large differential SA for speed in normal SRAM mode and separate into two single-ended SA for area efficiency in computation mode. Through a NOR gate, we can get $A\; \oplus \; B$ which is then used to generate the sum ($A\; \oplus\; B\; \oplus\; C_{in}$) and Carry (($A\;\&\; B$) + ($A \; \oplus\;B\; \& \; C_{in}$)). As described in the previous sections, $C$ and $T$ are latches used to store carry and tag bit. A 4-to-1 mux selects the data to be written back among $Sum$, $Carry_{out}$, $Data_{in}$, and $Tag$. The Tag bit is used as the enable signal for the bit line driver to decide whether to write back or not.


\subsection{Transpose Gateway Units}
\label{sec:transpose}
The \emph{transpose} data layout can be realized in the following
ways. First, leverage programmer support to store and access data in the
\emph{transpose} format. This option is useful when the data to be
operated on does not change at runtime. We utilize this for filter weights in neural networks. 
However, this approach
increases software complexity by requiring
programmers to reason about a new
data format and cache geometry. 

\begin{figure}[h]
	\minipage{0.5\textwidth}
\centering
  	\includegraphics[height=1.5in]{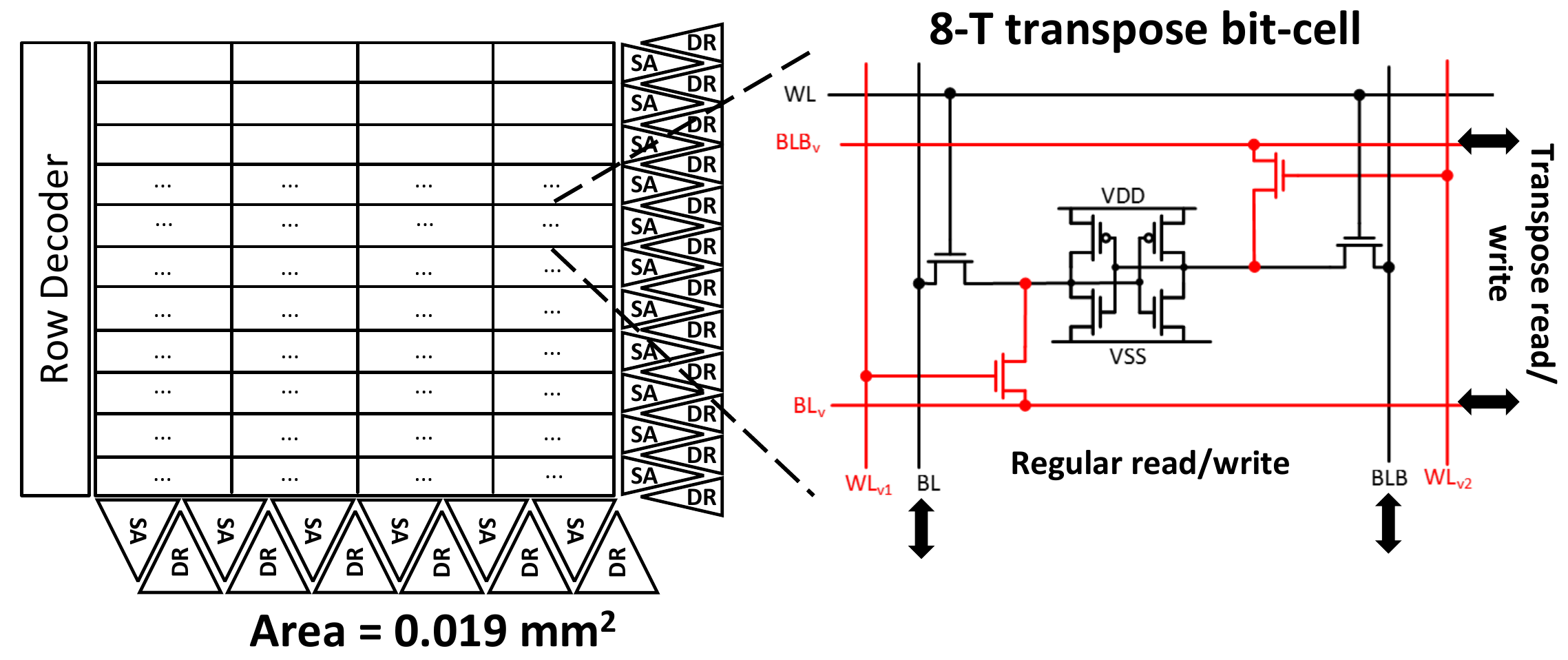}
	\endminipage
  \caption{Transpose Memory Unit (TMU)} 
\label{fig:tmu}
  \vspace{2mm}
\end{figure}
Second, design a few hardware \emph{transpose memory
  units (TMUs)} placed in the cache control box (C-BOX in Figure~\ref{fig:cc}~(b)). A TMU takes data in
the \emph{bit-parallel or regular} layout and converts it to the \emph{transpose}
layout before storing into SRAM arrays or vice-versa while reading
from SRAM arrays. The second option is
attractive because it supports dynamic changes to data. TMUs can be built
out of SRAM arrays with multi-dimensional access (i.e., access data in
both horizontal and vertical direction). Figure~\ref{fig:tmu} shows a
possible TMU design  using an 8T SRAM array with sense-amps in both
horizontal and vertical directions. Compared to a baseline 6T SRAM, the transposable SRAM requires a 
larger bitcell to enable read/write in both directions. Note that only a few TMUs are needed to saturate the
available interconnect bandwidth between cache arrays. In essence, the
transpose unit serves as a gateway to enable bit-serial computation in
caches.

\begin{figure*}
  \includegraphics[width=\textwidth]{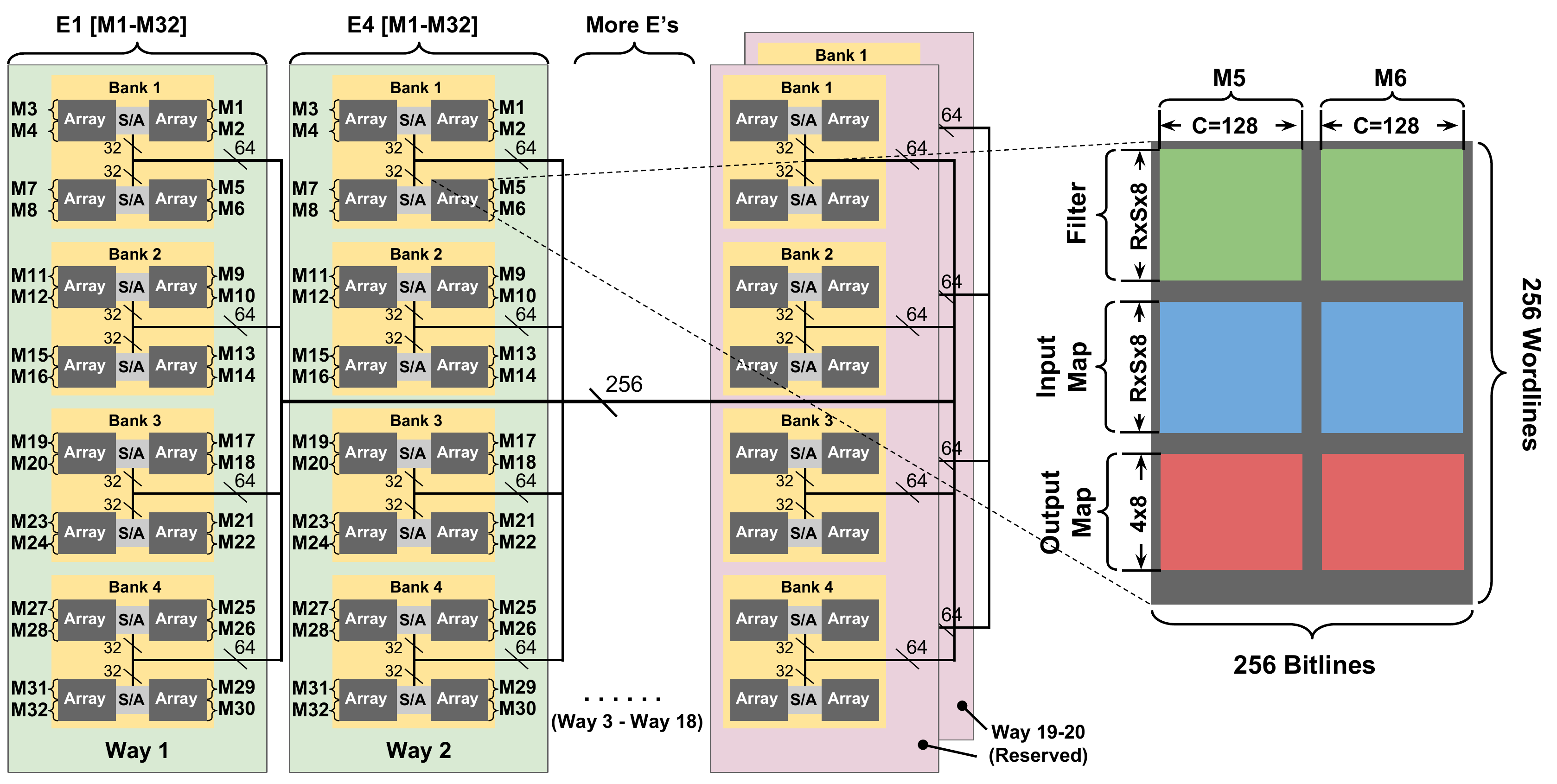}
  \caption{Neural Cache Data Layout for one LLC Cache Slice.}
  \label{fig:slicedm}
\end{figure*}

\section{Neural Cache Architecture}
\label{sec:arch}
The \nc{} architecture transforms SRAM arrays in LLC to compute functional units. We describe the computation of convolution layers first, followed by other layers. Figure~\ref{fig:slicedm} shows the data layout and overall architecture for one cache slice, modeled after Xeon processors~\cite{xeonllc45nm,xeonllc22nm}. The slice has twenty ways. The last way (way-20) is reserved to enable normal processing for CPU cores. The penultimate way (way-19) is reserved to store inputs and outputs. The remaining ways are utilized for storing filter weights and computing.

A typical DNN model consists of several layers, and each layer consists of several hundred thousands of convolutions. For example, Google's Inception v3 has 20 layers, most of which have several branches. Inception v3 has $\approx$ 0.5 million convolutions in each layer on average. \nc{} computes layers and each branches within a layer serially. The convolutions within a branch are computed in parallel to the extent possible. Each of the 8KB SRAM arrays computes convolutions in parallel. The inputs are streamed in from the reserved way-19. Filter weights are stationary in compute arrays (way-1 to way-18). 

\nc{} assumes 8-bit precision and quantized inputs and filter weights. Several works~\cite{gupta2015deep,han2015deep,zhou2016dorefa} have shown that 8-bit precision has sufficient accuracy for DNN inference. 8-bit precision was adopted by Google's TPU~\cite{jouppi2017datacenter}. Quantizing input data requires re-quantization after each layer as discussed in Section~\ref{sec:quant}.

\subsection{Data Layout}
\label{sec:dm}

This section first describes the data layout of one SRAM array and execution of one convolution. Then we discuss the data layout for the whole slice and parallel convolutions across arrays and slices. 

A \emph{single convolution} consists of generating \emph{one} of the $E \times E \times M$ output elements. This is accomplished by multiplying $R \times S \times C$ input filter weights with a same size window from the input feature map across the channels. \nc{} exploits channel level parallelism in a single convolution. For each convolution, an array executes $R \times S$ Multiply and Accumulate (MAC) in parallel across channels. This is followed by a reduction step across channels. 

\begin{figure}[h]
	\minipage{0.5\textwidth}
\centering
  	\includegraphics[scale=0.35]{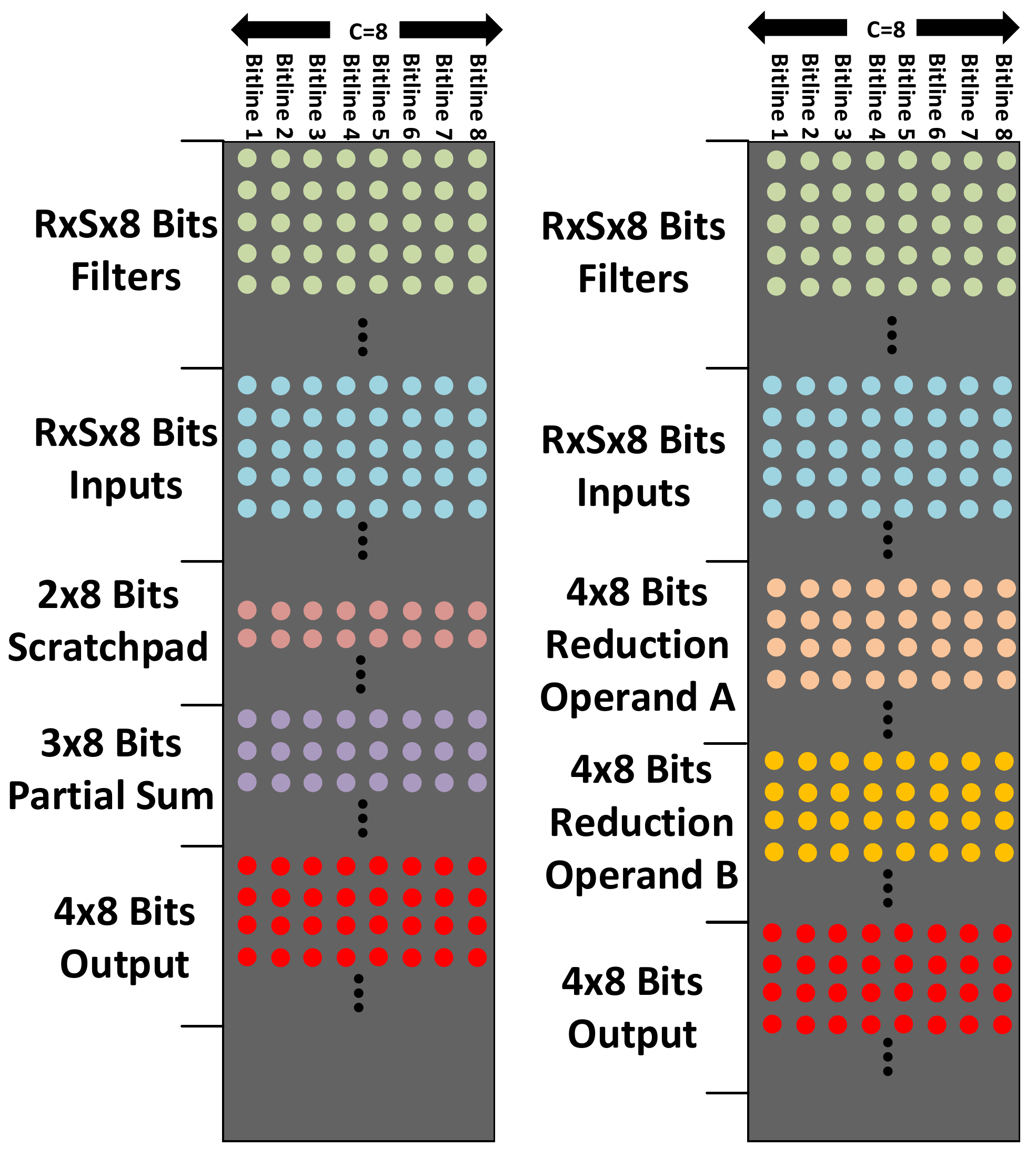}
	\endminipage
  \caption{Array Data Layout (a) MACs (b) Reduction.} 
	\label{fig:dm}
\end{figure}
An example layout for a  single array is shown in Figure~\ref{fig:dm} (a). Every bitline in the array has 256 bits and  can store 32 1-byte elements in transpose layout.  Every bitline stores $R \times S$ filter weights (green dots). The channels are stored across bit lines. To perform MACs, space is reserved for accumulating partial sums (lavender dots) and for scratch pad (pink dots). Partial sums and scratch pad take $3\times8$  and $2\times8$ word lines. 

Reduction requires an additional $8\times8$ word lines as shown in Figure~\ref{fig:dm} (b). However the scratch pad and partial sum can be overwritten for reduction as the values are no longer needed. The maximum size for reducing all partial sums is 4 bytes. So to perform reduction, we reserve two 4 byte segments. After adding the two segments, the resultant can be written over the first segment again. The second segment is then loaded with the next set of reduced data.  

Each array may perform several convolutions in series, thus we reserve some extra space for output elements (red dots). The remaining word lines are used to store input elements (blue dots). It is desirable to use as many word lines as possible for inputs to maximize input reuse across convolutions. For example in a $3\times3$ convolution with a stride of 1, 6 of the 9 bytes are reused across each set of input loads. Storing many input elements allows us to exploit this locality and reduce input streaming time. 

The filter sizes ($R \times S$) range from 1-25 bytes in Inception v3. The common case is a $3\times3$ filter. \nc{} data mapping employs \emph{filter splitting} for large filters. The filters are split across bitlines when their size exceeds 9 bytes. The other technique employed is \emph{filter packing}. For $1\times1$ filters we compress multiple channels into the same bit line. Instead of putting a single byte of the filter, we can instead put 16 bytes of the filter. Since $1\times1$ has no input reuse, we only need one input byte at a time. By packing the filters, the number of reductions is decreased at no cost to input loading. More importantly, by packing all channels in the network it is guaranteed to fit within 2 arrays that share sense-amps, making reduction faster and easier. 

When mapping the layers, first the new effective channel size after packing and filter splitting is calculated. This channel number is then rounded up to the nearest power of 2, by padding the extra channels with zero. Depending on parameters, filters from different output channels (M's) can be placed in the same array as well. For instance, the first layer of Inception v3 has so few channels, that all M's can be placed in the same array. 

Finally, although all operations are accomplished at the bit level, each data element is stored as a multiple of a byte, although it may not necessarily require that many bits. This is done for simplicity, software programmability, and easier data movement.
\subsection{Data Parallel Convolutions}
Each layer in the network produces $M\times E \times E$ outputs, and each output requires one convolution. All the outputs can be produced in parallel, provided we have sufficient computing resources. We find that in most scenarios half an array or even a quarter of an array is sufficient to computes each output element. Thus \nc{}'s massively parallel computing resources can be utilized to do convolutions in parallel. 

\begin{figure}[h]
	\minipage{0.5\textwidth}
\centering
  	\includegraphics[scale=0.42]{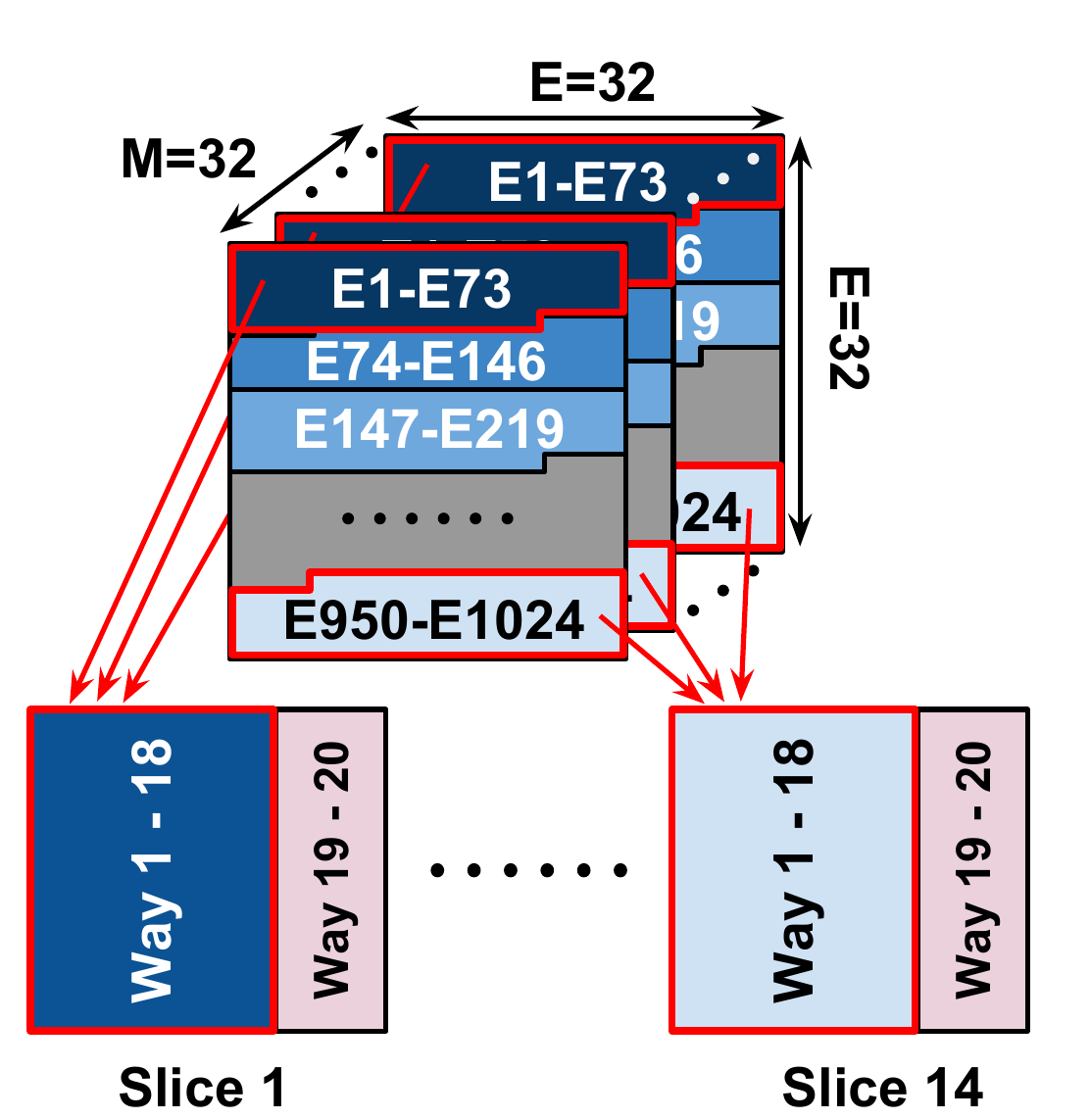}
	\endminipage
  \caption{Partitioning of Convolutions between Slices.} 
	\label{fig:workdiv}
\end{figure}
Figure~\ref{fig:workdiv} shows the division of convolutions among cache slices. Each slice can be visualized as a massively parallel SIMD core. Each slice gets roughly an equal number of outputs to work on. Mapping consecutive output elements to the same slice has the nice property that the next layer can find much of its required input data locally within its reserved way. This reduces the number of inter-slice transfers needed when loading input elements across layers.

Figure~\ref{fig:slicedm} shows the layout of one slice for an example layer. This layer has parameters: $R\times S=3\times3$, $C=128$, $M=32$, $E=32$. Figure~\ref{fig:slicedm} (right) shows a single array. Each array can pack two complete filters ($R\times S \times C$). The array shown in figure ~\ref{fig:slicedm} packs $M=5$ and $M=6$ in the same array. Thus each array can compute two convolutions in parallel. Each way (4 banks, or 16 arrays) computes 32 convolutions in parallel ($E_{i}\; \forall\; M's\; M1-M32$). The entire slice can compute $18 \times 32$ convolutions in parallel. Despite this parallelism, some convolutions need to be computed serially, when the total number of convolutions to be performed exceed the total number of convolutions across \emph{all slices}. In this example, about 4 convolutions are executed serially. 

When mapping inputs and filters to the cache, we prioritize uniformity across the cache over complete (i.e., 100\%) cache compute resource utilization. This simplifies operations as all data is mapped in such a way that all arrays in the cache will be performing the exact same instruction at the same time and also simplifies data movement.

\subsection{Orchestrating Data Movement}

Data parallel convolutions require replication of filters and streaming of inputs from the reserved way. This section discusses these data-movement techniques. 

We first describe the interconnect structure. Our design is based on the on-chip LLC \cite{xeonllc45nm} of the Intel Xeon E5-2697 processor. There are in total 14 cache slices connected by a bidirectional ring. Figure~\ref{fig:slicedm} shows the interconnect within a cache slice. There is a 256-bit data bus which is capable of delivering data to each of the 20 ways. The data bus is composed of 4 64-bit data buses. Each bus serves one quadrant. A quadrant consists of a 32KB bank composed of four 8KB data arrays. Two 8 KB arrays within a bank share sense-amps and receive 32 bits every bus cycle. 

\textbf{Filter Loading:} We assume that all the filter data for a specific layer reside in DRAM before the entire computation starts. Filter data for different layers of the DNN model are loaded in serial. Within a convolution layer, regardless of the number of output pixels done in serial, the positions of filter data in the cache remain stationary. Therefore, it is sufficient to load filter data from \emph{memory only once per layer}.

 Across a layer we use M sets of $R\times S$ filters. By performing more than M convolutions, we will replicate filters across and within slices.
Fortunately the inter-slice ring and intra-slice bus are both naturally capable of broadcasting, allowing for easy replication for filter weights. Each filter weight loaded from DRAM is broadcasted to all slices over the ring and all ways over the intra-slice bus. 
 
 In each array that actively performs computation, $R'\times S'\times 8$ word lines are loaded with filter data, where $R'\times S'$ is the equivalent filter dimension after packing and filter splitting, and 8 is the bit-serial factor due to 8-bits per element. \nc{} assumes that filter weights are preprocessed to a transpose format and laid out in DRAM such that they map to correct bitlines and word-lines. Our experiments decode the set address and faithfully model this layout. Software transposing of weights is a one time cost and can be done cheaply using x86 SIMD shuffle and pack instructions~\cite{parabixtrans,parabix}.


\textbf{Input Data Streaming:} We only load the input data of the first layer from DRAM, because for the following layers, the input data have already been computed and temporarily stored in the cache as outputs. In some layers, there are multiple output pixels to be computed in serial, and we need to stream in the corresponding input data as well, since different input data is required at any specific cache array for generating different output pixels.

Within each array that actively performs computation, $R'\times S' \times 8$ word lines of input data are streamed in, where $R'\times S'$ is the equivalent filter dimension after packing and filter splitting, and 8 is the bit-serial factor. When loading inputs from DRAM for the first layer, input data is transposed using the TMUs at C-BOX. 

Since, each output pixel requires a different set of input data, input loading time can be high. We exploit duplicity in input data to reduce transfer time. For different output channels (M's) with the same output pixel position (i.e. same $E_{i}$ for different M's), the input data is the same and can be replicated. Thus for scenarios were these different output channels are in different ways, the input data can be loaded in one intra-slice bus transfer. Furthermore, a large portion of input data can be reused across output pixels done in serial as discussed in Section~\ref{sec:dm}. This helps reducing transfer time. We also observe that often the input data is replicated across arrays within a bank. We put a 64-bit latch at each bank, so that the total input transfer time can be halved. 

Note, intra-bus transfers happen in parallel across different cache slices. Thus distributing E's across slices significantly reduces input loading time as well by leveraging intra-slice network bandwidth.  

\textbf{Output Data Management:} One way of each slice (128 KB), is reserved for temporarily storing the output data. After all the convolutions for a layer are executed, data from compute arrays are transferred to the reserved way. We divide the computation into different slices according to the output pixel position. Contiguous output pixels are assigned to the same slice so that one slice will need neighboring inputs for at most $R\times E$ pixels. This design significantly reduces inter-slice communication for transferring outputs.

\subsection{Supporting Functions}
\label{sec:quant}

\textbf{Max Pooling} layers compute the maximum value of all the inputs inside a sliding window. The data mapping and input loading would function the same way as convolution layers, except without any filters in the arrays.

Calculating the maximum value of two or more numbers can be accomplished by designating a temporary maximum value.  The temporary maximum is then subtracted by the next output value and the resultant is stored in a separate set of word lines. The most significant bit of the result is used as a mask for a selective copy. The next input is then selectively copied to the maximum location based on the value of the mask. This process is repeated for the rest of the inputs in the array. 

\textbf{Quantization} of the outputs is done by calculating the the minimum and maximum value of all the outputs in the given layer. The min can be computed using a similar set of operations described for max. For quantization, the min and max will first be calculated within each array. Initially all outputs in the array will be copied to allow performing the min and max at the same time.  After the first reduction, all subsequent reductions of min/max are performed the same way as channel reductions. Since quantization needs the min and max of the entire cache, a series of bus transfers is needed to reduce min and max to one value.  This is slower than in-array reductions, however unlike channel reduction, min/max reduction happens only once in a layer making the penalty small. 

After calculating the min and max for the entire layer, the result is then sent to the CPU. The CPU then performs floating point operations on the min and max of the entire layer and computes two unsigned integers. These operations take too few cycles to show up in our profiling. Therefore, it is assumed to be negligible. The two unsigned integers sent back by the CPU are used for in-cache multiplications, adds, and shifts to be performed on all the output elements to finally quantize them.


\textbf{Batch Normalization} requires first quantizing to 32 bit unsigned. This is accomplished by multiplying all values by a scalar from the CPU and performing a shift. Afterwards scalar integers are added to each output in the corresponding output channel. These scalar integers are once again calculated in the CPU. Afterwards, the data is re-quantized as described above. 

In Inception v3, \textbf{ReLU} operates  by replacing any negative number with zero. We can write zero to every output element with the MSB acting as an enable for the write. Similar to max/min computations, ReLU relies on using a mask to enable selective write.

\textbf{Avg Pool} is mapped in the same way as max pool. All the inputs in a window are summed and then divided by the window size. Division is slower than multiplication, but the divisor is only 4 bits in Inception v3.

\textbf{Fully Connected} layers are converted into convolution layers in TensorFlow. Thus, we are able to treat the fully connected layer as another convolution layer. 

\subsection{Batching}
We apply batching to increase the system throughput. Our experiments show that loading filter weights takes up about 46\% of the total execution time. Batching multiple input images significantly amortizes the time for loading weights and therefore increases system throughput. \nc{} performs batching in a straightforward way. The image batch will be processed sequentially in the layer order. For each layer, at first, the filter weights are loaded into the cache as described in Section~\ref{sec:dm}. Then, a batch of input images are streamed into the cache and computation is performed in the same way as without batching. For the whole batch, the filter weights of the involved layer remain in the arrays, without reloading. Note that for the layers with heavy-sized outputs, after batching, the total output size may exceed the capacity of the reserved way. In this case, the output data is dumped to DRAM and then loaded again into the cache. In the Inception v3, the first five requires dumping output data to DRAM.

\subsection{ISA support and Execution Model}

\nc{} requires supporting a few new instructions: in-cache addition, multiplication, reduction, and moves. Since, at any given time only one layer in the network is being operated on, all compute arrays execute the same in-cache compute instruction. The compute instructions are followed by move instructions for data management. The intra-slice address bus is used to broadcast the instructions to all banks. Each bank has a control FSM which orchestrates the control signals to the SRAM arrays. The area of one FSM is estimated to be 204 $\mu m^{2}$, across 14 slices which sums to 0.23 $mm^{2}$. Given that each bank is executing the same instruction, the control FSM can be shared across a way or even a slice. We chose not to optimize this because of the insignificant area overheads of the control FSM. \nc{} computation is carried out in 1-19 ways of each slice. The remaining way (way-20) can be used by other processes/VMs executing on the CPU cores for normal background operation. Intel's Cache Allocation Technology (CAT)~\cite{intelcat} can be leveraged to dynamically restrict the ways accessed by CPU programs to the reserved way.

\begin{figure*}[thb]
   \centering
	  	\includegraphics[scale=0.5] {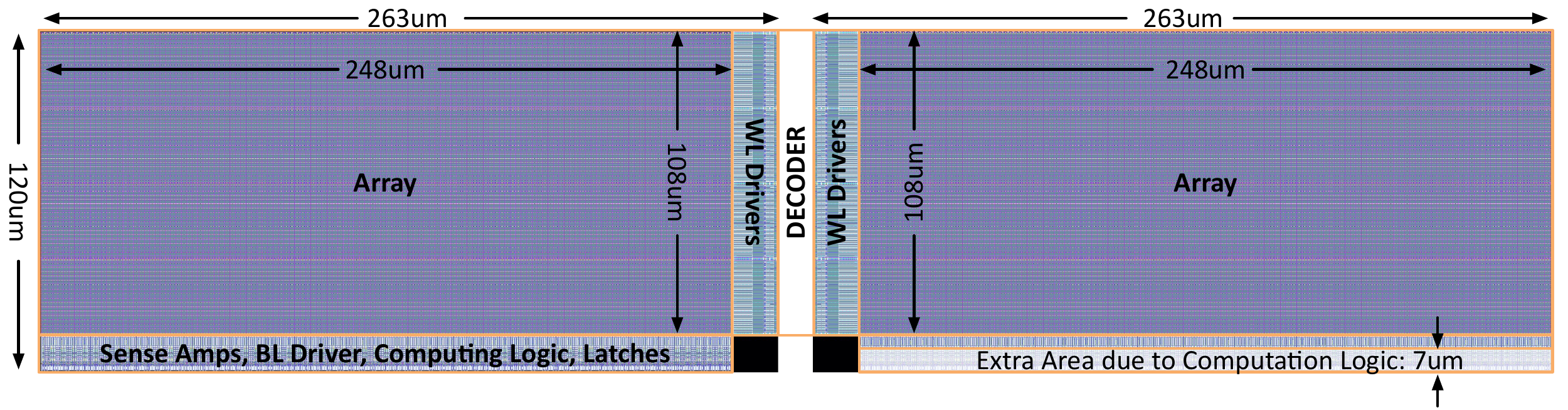}
	  \caption{SRAM Array Layout.} 
	\label{fig:sraml}
    \vspace{4mm}
\end{figure*}
\begin{table*}[thb]
\small
\begin{center}
\begin{tabular}{| c | c | c | c | c | c | c | c | c |}
\hline
Layer & H & $R \times S$ & E & C & M & Conv & Filter Size / MB & Input Size / MB \\ \hline
Conv2D\_1a\_3x3 & 299 & 9 & 149 & 3 & 32 & 710432 & 0.001 & 0.256\\ \hline
Conv2D\_2a\_3x3 & 149 & 9 & 147 & 32 & 32 & 691488 & 0.009 & 0.678\\ \hline
Conv2D\_2b\_3x3 & 147 & 9 & 147 & 32 & 64 & 1382976 & 0.018 & 0.659\\ \hline
MaxPool\_3a\_3x3 & 147 & 9 & 73 & 0 & 64 & 0 & 0.000 & 1.319\\ \hline
Conv2D\_3b\_1x1 & 73 & 1 & 73 & 64 & 80 & 426320 & 0.005 & 0.325\\ \hline
Conv2D\_4a\_3x3 & 73 & 9 & 71 & 80 & 192 & 967872 & 0.132 & 0.407\\ \hline
MaxPool\_5a\_3x3 & 71 & 9 & 35 & 0 & 192 & 0 & 0.000 & 0.923\\ \hline
Mixed\_5b & 35 & 1-25 & 35 & 48-192 & 32-192 & 568400 & 0.243 & 0.897\\ \hline
Mixed\_5c & 35 & 1-25 & 35 & 48-256 & 48-256 & 607600 & 0.264 & 1.196\\ \hline
Mixed\_5d & 35 & 1-25 & 35 & 48-288 & 48-288 & 607600 & 0.271 & 1.346\\ \hline
Mixed\_6a & 35 & 1-9 & 17 & 64-288 & 64-384 & 334720 & 0.255 & 1.009\\ \hline
Mixed\_6b & 17 & 1-9 & 17 & 128-768 & 128-768 & 443904 & 1.234 & 0.847\\ \hline
Mixed\_6c & 17 & 1-9 & 17 & 160-768 & 160-768 & 499392 & 1.609 & 0.847\\ \hline
Mixed\_6d & 17 & 1-9 & 17 & 160-768 & 160-768 & 499392 & 1.609 & 0.847\\ \hline
Mixed\_6e & 17 & 1-9 & 17 & 192-768 & 192-768 & 499392 & 1.898 & 0.847\\ \hline
Mixed\_7a & 17 & 1-9 & 8 & 192-768 & 192-768 & 254720 & 1.617 & 0.635\\ \hline
Mixed\_7b & 8 & 1-9 & 8 & 384-1280 & 192-1280 & 208896 & 4.805 & 0.313\\ \hline
Mixed\_7c & 8 & 1-9 & 8 & 384-2048 & 192-2048 & 208896 & 5.789 & 0.500\\ \hline
AvgPool & 8 & 64 & 1 & 0 & 2048 & 0 & 0.000 & 0.125\\ \hline
FullyConnected & 1 & 1 & 1 & 2048 & 1001 & 1001 & 1.955 & 0.002\\ \hline
\end{tabular}
\caption{Parameters of the Layers of Inception V3.}
\label{table:incept_param}
\end{center}
\end{table*}
\section{Evaluation Methodology}

\begin{table}
\small
\begin{center}
\begin{tabular}{| c | c |}
\hline
\textbf{CPU} & \textbf{Intel Xeon E5-2697 v3}  \\ \hline
Base Frequency & 2.6 GHz\\\hline
Cores/Threads & 14/28\\\hline
Process & 22 nm\\ \hline
TDP & 145 W\\\hline
Cache & \begin{tabular}[t]{@{}c@{}}32 kB i-L1 per core, 32 kB d-L1 per core,\\256 kB L2 per core, 35 MB shared L3\end{tabular} \\ \hline
System Memory & 64 GB DRAM, DDR4\\ \hline

\end{tabular}
\\ \vspace{1mm}
\begin{tabular}{| c | c |}
\hline
\textbf{GPU} &  \textbf{Nvidia Titan Xp} \\ \hline
Frequency & 1.6 GHz\\\hline
CUDA Cores & 3840 \\ \hline
Process & 16 nm \\ \hline
TDP & 250 W \\ \hline
Cache & 3MB shared L2 \\\hline
Graphics Memory & 12 GB DRAM, GDDR5X\\ \hline
\end{tabular}
\caption{Baseline CPU \& GPU Configuration.}
\label{table:baseconf}
\vspace{-8mm}
\end{center}
\end{table}

\textbf{Baseline Setup:} For baseline, we use dual-socket Intel Xeon E5-2697 v3 as CPU, and Nvidia Titan Xp as GPU. The specifications of the baseline machine are in Table \ref{table:baseconf}. Note that the CPU specs are per socket. Note that the baseline CPU has the exact cache organization (35 MB L3 per socket) as we used in \nc{} modeling.
The benchmark program is the inference phase of the Inception v3 model \cite{szegedy2016rethinking}. We use TensorFlow as the software framework to run NN inferences on both baseline CPU and GPU. The default profiling tool of TensorFlow is used for generating execution time breakdown by network layers for both CPU and GPU. The reported baseline results are based on the unquantized version of Inception v3 model, because we observe that the 8-bit quantized version has a higher latency on the baseline CPU due to lack of optimized libraries for quantized operations (540 ms for quantized / 86 ms for unquantized).
To measure execution power of the baseline, we use RAPL\cite{david2010rapl} for CPU power measurement and Nvidia-SMI \cite{nvidiasmi} for GPU power measurement.

\textbf{Modeling of Neural Cache:} To estimate the power and delay of the SRAM array, the SPICE model of an 8KB
computational SRAM is simulated using the 28 $nm$ technology node. The nominal voltage for this
technology is 0.9V. Since we activate two read word lines (RWL) at the same time in
computation, we reduce the RWL voltage to improve bit cell read stability. But lowering RWL voltage will
increase the read delay. We simulated for various RWL voltages and to achieve industry standard 6 sigma margin, we choose
0.66V as the RWL voltage. The total computing time is 1022ps. The delay of a normal SRAM
read is 654ps given by the standard foundry memory compiler. So the
computation SRAM delay is about $1.6\times$ larger than normal SRAM read. Xeon processor cache arrays can operate at 4 GHz~\cite{xeonllc45nm,xeonllc22nm}. We conservatively choose a frequency of 2.5 GHz for \nc{} in the compute mode.
Our SPICE simulations also provided the
total energy consumption in one clock cycle for reading out 256 bits in SRAM mode or operating on
256 bit lines in computation mode. This was estimated to be 13.9pJ for SRAM access cycles and 25.7pJ for compute cycles. Since we model \nc{} for the Intel Xeon E5-2697 v3 processor at 22 $nm$, the energy was scaled down to 8.6pJ for SRAM access cycles and 15.4pJ for compute cycles. The SRAM array layout is shown in Figure~\ref{fig:sraml}. Compute capabilities incur an area overhead of 7.5\% increase due to extra logic  and an extra decoder.

For the in-cache computation, we developed a cycle-accurate simulator based on the deterministic computation model discussed in Section~\ref{sec:arch}. The simulator is verified by running data traces on it and matching the results with traces obtained from instrumenting the TensorFlow model for Inception v3. To model the time of data loading, we write a C micro-benchmark, which sequentially reads out the exact sets within a way that need loading with data. The set decoding was reverse engineered based on Intel's last level cache architecture~\cite{xeonllc45nm,xeonllc22nm}. We conduct this measurement for all the layers according to their different requirement of sets to be loaded. Then, the measured time is multiplied with a factor that accounts for the sequential data transfer across different ways, as well as the sequential computation of different output pixels within a layer. Note that the measured time includes the data loading from DRAM to on-chip cache, but for input data loading, the data is already in the cache (except the first layer). For more accurate modeling, the micro-benchmark is profiled by the VTune Amplifier\cite{intelvtune} for estimating the percentage of DRAM bound cycles, and such DRAM-loading time is excluded from the input data loading time.

\begin{figure}[t]
 \begin{minipage}{0.5\textwidth}
  \begin{center}
		\includegraphics[scale=.38]{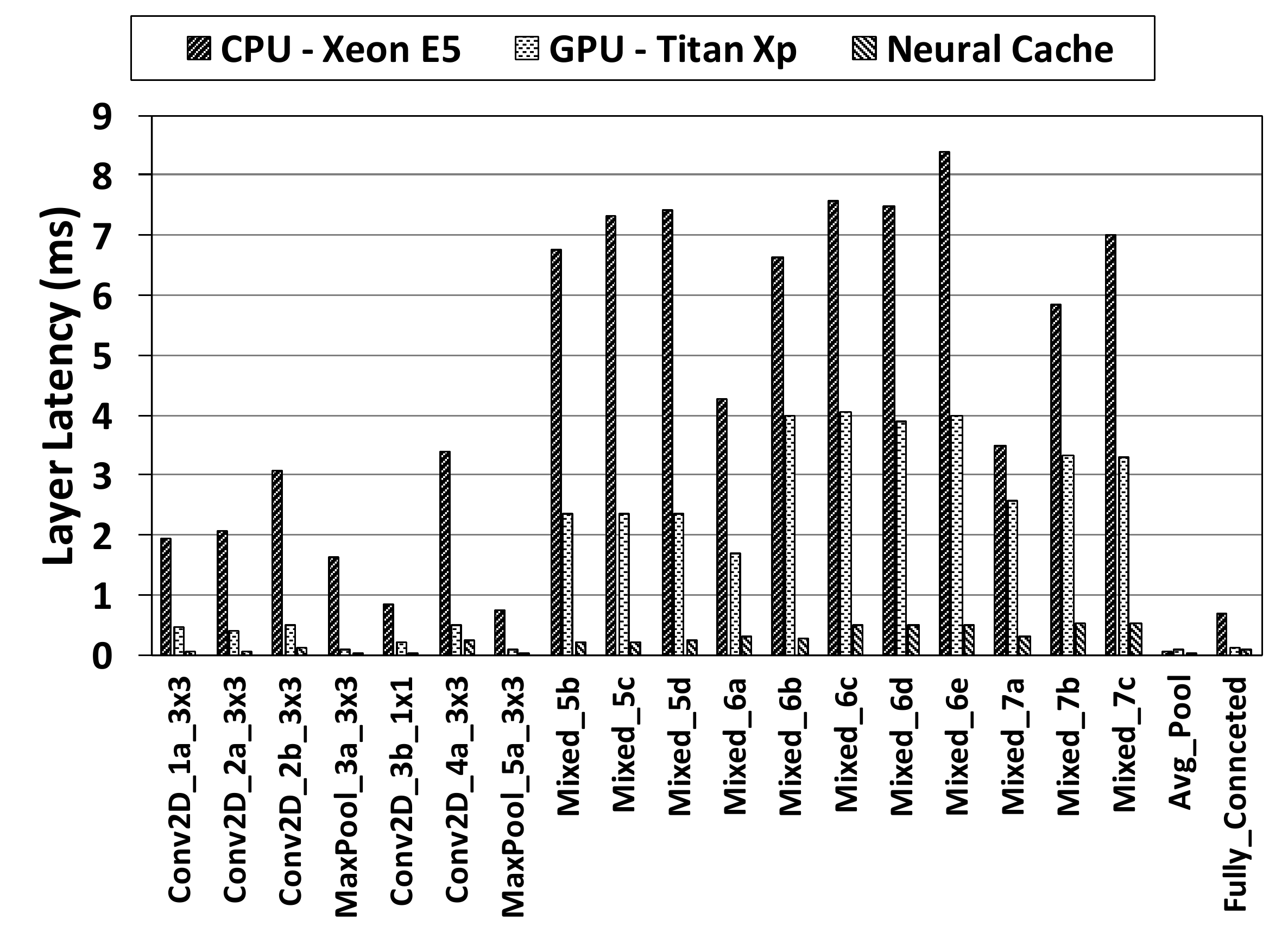}
  \end{center}
  \caption{Inference latency by Layer of Inception v3.}
		\label{fig:barlayer}
         \vspace{3mm}
 \end{minipage}
 \begin{minipage}{0.5\textwidth}
  \begin{center}
       \includegraphics[scale=0.38]{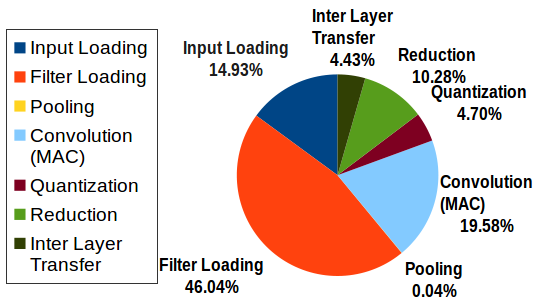}
  \end{center}
  \caption{Inference Latency Breakdown.}
		\label{fig:piechart}
 \end{minipage}
\end{figure}
\section{Results}
\vspace{-3mm}
\subsection{Latency}
 Figure \ref{fig:barlayer} reports the latency of all layers in the Inception v3 model. A majority of time is spent on the mixed layers for both CPU and GPU, while \nc{} achieves significantly better latency than baseline for all layers. This is primarily because \nc{} exploits the available data parallelism across convolutions to provide low-latency inference. \nc{}'s data mapping not only makes the computation entirely data independent, but the operations performed are identical, allowing for SIMD instructions. 
 
 Consider an example layer, Conv2D\_Layer\_2b\_3$\times$3. This layer computes $\approx 1.4\;million$ convolutions, out of which \nc{} executes $\approx 32\;thousand$  convolutions in parallel and 43 in series. The compute cache arrays show 99.7\% utilization for this layer during convolutions(after data loading). Each convolution takes 2784 cycles (236 cycles/MAC$\times$ 9 + 660 reduction cycles). Then reduction takes a further 660 cycles. The whole layer takes 117912 cycles (43 convolution in series $\times$ 2784 cycles), taking 0.0479 ms to finish the convolutions for \nc{} running at 2.5 GHz. Remaining time for the layer is attributed to data loading. CPU and GPU cannot take advantage of data parallelism on this scale due to lack of sufficient compute resources and on-chip data-movement bottlenecks.

 Figure \ref{fig:piechart} shows the breakdown of \nc{} latency. Loading filter weights into cache and distributing them into arrays takes up 46\% of total execution time, and 15\% of the time is spent on streaming in the input data to the appropriate position within the cache. Transferring output data to the proper reserved space takes 4\% of the time. Filter loading is the most time consuming part since data is loaded from DRAM. The remaining parts are for computation, including multiplication in convolution (MACs) (20\%), reduction (10\%), quantization (5\%), and pooling (0.04\%).
 
\begin{figure}[h]
	\minipage{0.5\textwidth}
\centering
  	\includegraphics[scale=0.25]{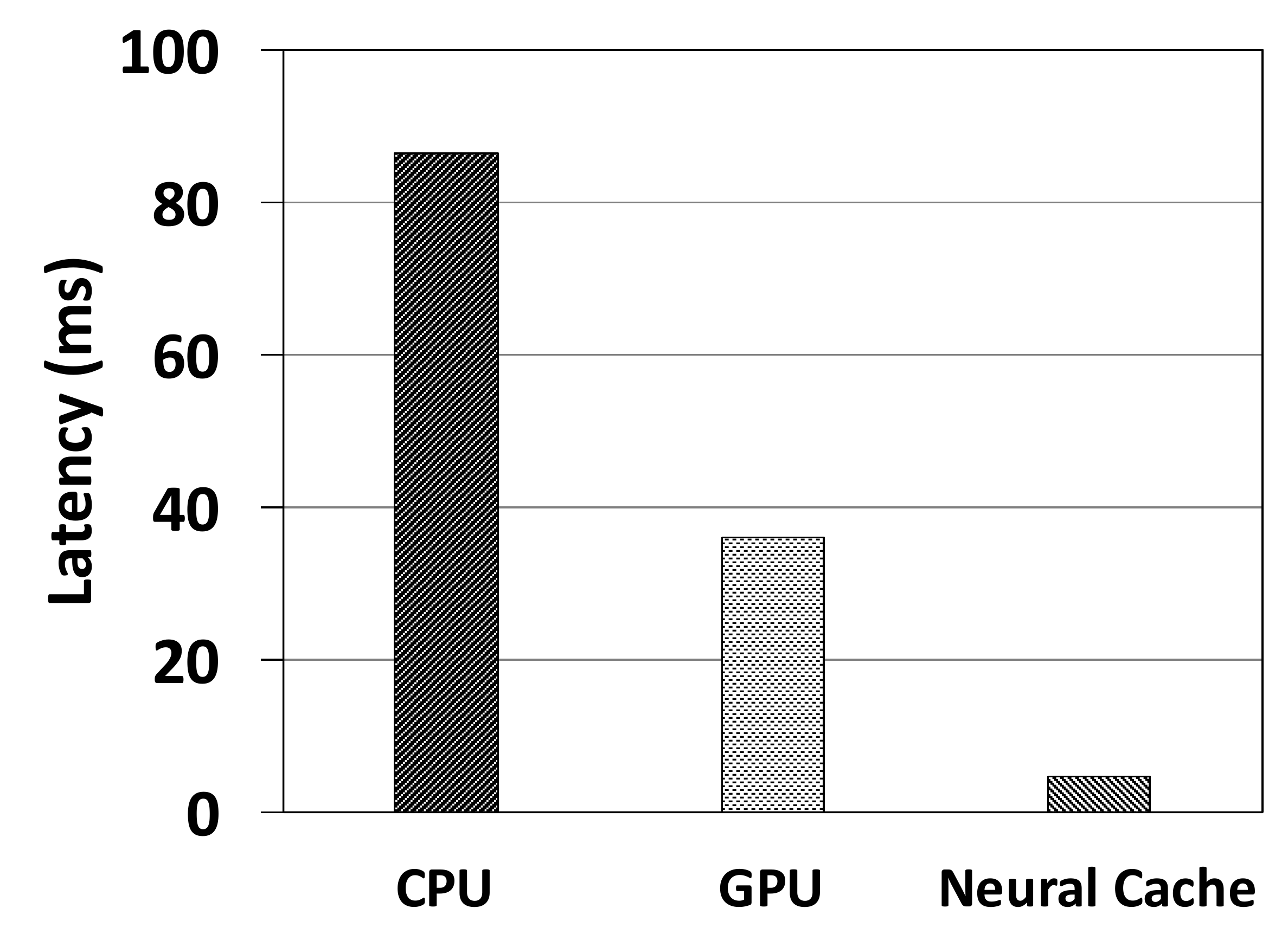}
	\endminipage
  \caption{Total Latency on Inception v3 Inference.} 
	\label{fig:totallat}
\end{figure}

 Figure \ref{fig:totallat} shows the total latency for \nc{}, as well as the baseline CPU and GPU, running inference on the Inception v3 model. \nc{} achieves a $7.7\times$ speedup in latency compared to baseline GPU, and $18.3\times$ speedup on baseline CPU. The significant speedup can be attributed to the elimination of high overheads of on-chip data movement from cache to the core registers, and data parallel convolutions.

\subsection{Batching}
\begin{figure}[h]
	\minipage{0.5\textwidth}
\centering
  	\includegraphics[scale=0.35]{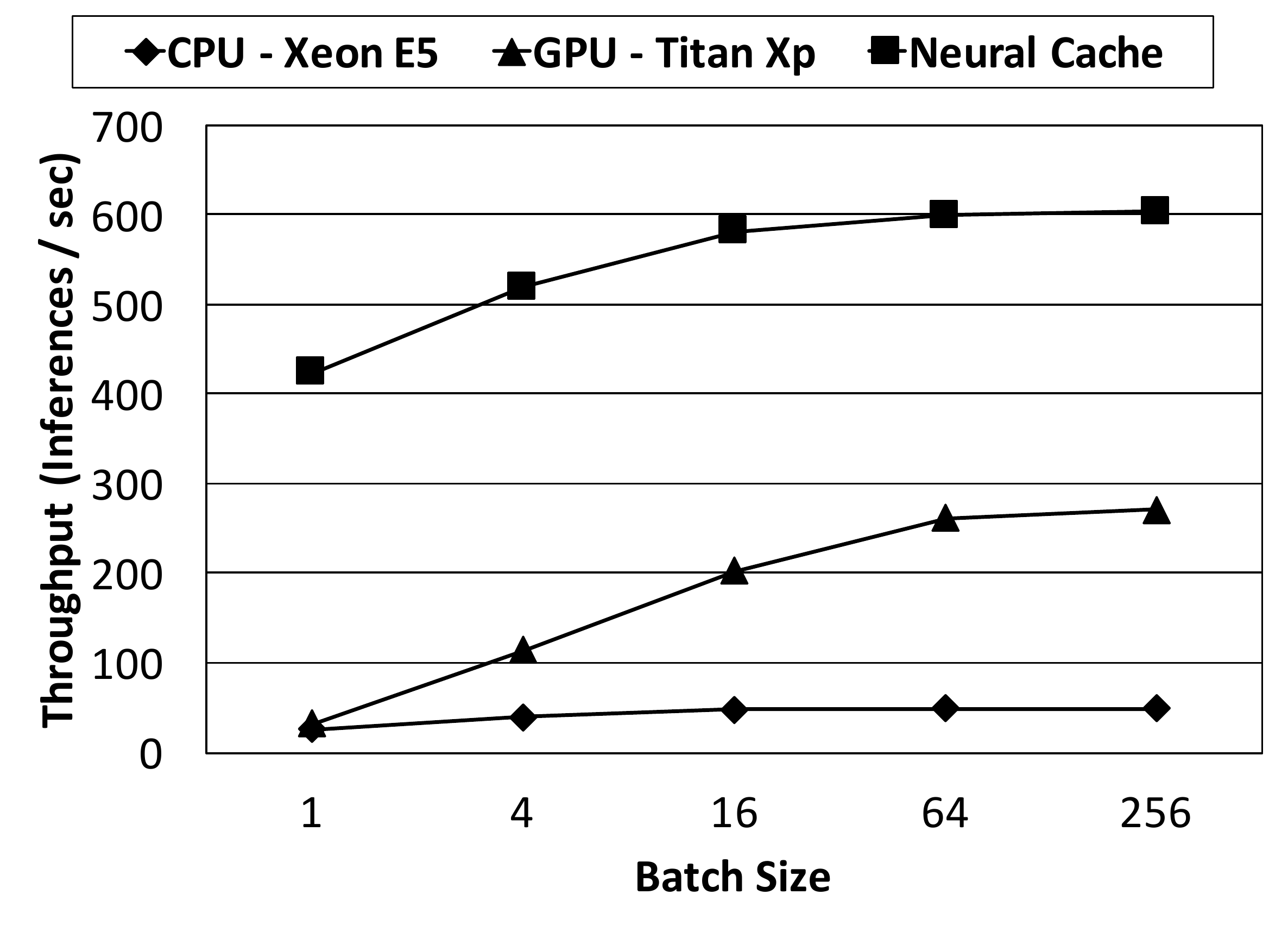}
	\endminipage
  \caption{Throughput with Varying Batching Sizes.} 
	\label{fig:throughput}
\end{figure}

Figure \ref{fig:throughput} shows the system throughput in number of inferences per second as the batch size varies. \nc{} outperforms the maximum throughput of baseline CPU and GPU even without batching. This is mainly due to the low latency of \nc{}. Another reason is that \nc{} scales linearly with the number of host CPUs, and thus the throughput of \nc{} doubles with a dual-socket node.

As the batch size $N$ increases from 1 to 16, the throughput of \nc{} increases steadily. This can be attributed to the amortization of filter loading time. For even higher batch sizes, the effect of such amortization diminishes and therefore the throughput plateaus. As shown in the figure, the GPU throughput also plateaus after batch size exceeds 64. At the highest batch size, \nc{} achieves a throughput of 604 inferences/sec, which is equivalent to $2.2\times$ GPU throughput, or $12.4\times$ CPU throughput.

\subsection{Power and Energy}
This section discusses the energy and power consumption of \nc{}. Table \ref{table:epret} summarizes the energy/power comparison with baseline. \nc{} achieves an energy efficiency that is $16.6\times$ better than the baseline GPU and $37.1\times$ better than CPU. The energy efficiency improvement can be explained by the reduction in data movement, the increased instruction efficiency of SIMD-like architecture, and the optimized in-cache arithmetic units.


The average power of \nc{} is 53.11\% and 49.87\% lower than the GPU and CPU baselines respectively. Thus, \nc{} does not form a thermal challenge for the servers. \nc{} also outperforms both CPU and GPU baselines in terms of energy-delay product (EDP), since it consumes less energy and has a shorter latency.

\begin{table}[h]
\begin{center}
\begin{tabular}{| c | c | c | c |}
\hline
 & \textbf{CPU} & \textbf{GPU} & \textbf{Neural Cache}  \\ \hline
Total Energy / J & 9.137 & 4.087 & 0.246  \\ \hline
Average Power / W & 105.56 & 112.87 & 52.92\\ \hline
\end{tabular}
\caption{Energy Consumption and Average Power.}
\label{table:epret}
\end{center}
\end{table}

\begin{table}[h]
\begin{center}
\begin{tabular}{| c | c | c | c |}
\hline
\textbf {Cache Capacity} & 35MB & 45MB & 60MB  \\ \hline
\textbf{Inference Latency} & 4.72 ms & 4.12 ms & 3.79 ms  \\ \hline
\end{tabular}
\caption{Scaling with Cache Capacity (Batch Size=1).}
\label{table:sccap}
\end{center}
\end{table}
\vspace{-2mm}
\subsection{Scaling with Cache Capacity}
In this section we explore how increasing the cache capacity affects the performance of \nc{}. We increase the cache size from 35MB (14 slices) to 45MB (18 slices), and 60MB (24 slices). Increasing the number of slices speedups most aspects of the inference. The total number of arrays which compute increases, thereby increasing convolutions that can be done in parallel. This reduces the convolution compute time. Filter loading will not be affected as unique filters are not done across slices, rather filters are replicated across the slices. Input loading, however, will decrease since we are using the additional intra-slice bandwidth of new slices to decrease the time it takes to broadcast inputs. Inter-layer data transfer will also be reduced due to less outputs being transfered in each slice.

\section{Related Work}
To the best of our knowledge, this is the first work that exploits in-situ SRAM computation for accelerating inferences on DNNs. Below we discuss closely related prior works. 

\textbf{In-Memory Computing:}
\nc{} has its roots in the processing-in-memory (PIM) line of work~\cite{Kogge94,Patterson97}. PIMs  move logic {\em near} main memory (DRAM), and thereby reduce the gap between memory and compute units. \nc{}, in contrast, morphs cache (SRAM) structures into compute units, keeping data {\em in-place}. 

It is unclear if it would be possible to perform true in-place DRAM operations. There are four challenges. First, DRAM writes are destructive, thus in-place computation will corrupt input operand data. Solutions which copy data and compute on them are possible~\cite{seshadri2017ambit}, but slow. Second, the margin for sensing data in DRAM bit-cells is small, making sensing of computed values error prone. Solutions which change DRAM bit-cell are possible but come with 2-3$\times$ higher area overheads~\cite{li2017drisa, seshadri2017ambit}. Third, bit-line peripheral logic needed for arithmetic computation has to be integrated with DRAM technology which is not optimized for logic computation. Finally, data scrambling and address scrambling in commodity DRAMs~\cite{gen6-datasheet,intel-me,sgx-me} make re-purposing commodity DRAM for computation and data layout challenging.

While in-place operations in Memristors is promising~\cite{shafiee2016isaac,chi2016prime}, Memristors remain a speculative technology and are also significantly slower than SRAM. 

As caches can be found in almost all modern processors, we envision \nc{} to be a disruptive technology that can  enhance  commodity processors with large data-parallel accelerators almost free of cost. CPU vendors (Intel, IBM, etc.) can thus continue to provide high-performance general-purpose processors, while enhancing them with a co-processor like capability to exploit massive data-parallelism. Such a processor design is particularly attractive for difficult-to-accelerate applications that frequently switch between sequential and data-parallel computation.

\textbf{ASICs and FPGAs:}
Previously, a variety of neural network ASIC accelerators have been proposed. We discuss a few below.  
 DaDianNao~\cite{chen2014dadiannao} is an architecture for DNNs which integrates filters into on-chip eDRAM. It relieves the limited memory bandwidth problem by minimizing external communication and doing near-memory computation. 
Eyeriss~\cite{chen2016eyeriss} is an energy-efficient DNN accelerator which reduces data movement by maximizing local data reuse. It leverages row-stationary data flow to adapt to different shapes of DNNs. It adopts a spatial architecture consisting of 168 PE arrays connected by a NoC. In Eyeriss terminology, \nc{} has a hybrid of weight stationary and output stationary reuse pattern. Neurocube~\cite{neurocube-isca16} is a 3D DRAM accelerator solution which requires additional specialized logic integrated memory chips while \nc{} utilizes widely-used SRAM. 
The Tensor Processing Unit (TPU)~\cite{jouppi2017datacenter} is another ASIC for accelerating DNN inferences. The TPU chip features 
a high-throughput systolic matrix multiply unit for 8-bit MAC, as well as 28 MB on-chip memory. 

 In general, custom ASIC accelerator solutions achieve high efficiency while requiring extra hardware and incurring design costs.
 ASICs lack flexibility in that they cannot be re-purposed for other domains.  In contrast, our work is based on the cache, which improves performance of many other workloads when not functioning as a DNN accelerator. \nc{} aims to achieve high performance, while allowing flexibility of general purpose processing. Further, \nc{} is limited by commercial SRAM technology and general purpose processor's interconnect architecture. A custom SRAM accelerator ASIC can potentially achieve significantly higher performance than \nc{}. Being a SRAM technology, we also expect the compute efficiency of \nc{} to improve with newer technology nodes. 

The BrainWave project~\cite{chung2017accelerating} builds an architecture consisting of FPGAs connected with a custom network, for providing accelerated DNN service at a datacenter scale. The FPGA boards are placed between network switches and host servers to increase utilization and reduce communication latency between FPGA boards. The FPGA board features a central matrix vector unit, and can be programmed with a C model with ISA extensions. BrainWave with a network of Stratix 10 280 FPGAs at 14 $nm$ is expected to have 90 TOPs/s, while \nc{} achieves 28 TOPs/s at 22 $nm$ technology without requiring any additional hardware. BrainWave with current generation FPGAs achieves 4.7 TOPs/s. 

Terasys presents a bit-serial arithmetic PIM architecture\cite{gokhale1995processing}. Terasys reads the data out and performs the compute in bit-serial ALU's outside the array. \nc{} differs by performing partial compute along the bitlines and augments it with a small periphery to perform arithmetic in an area efficient architecture. Further Terasys performs software transposes while \nc{} has a dedicated hardware transpose unit, the TMU.

Bit-serial computation exploits parallelism at the level of numerical representation. Stripes~\cite{judd2016stripes} leverages bit-serial computation for inner product calculation to accelerate DNNs. Its execution time scales proportionally with the bit length, and thus enables a direct trade-off between precision and speed.
Our work differs from Stripe in that \nc{} performs in-situ computation on SRAM cells, while Stripe requires arithmetic functional units and dedicated eDRAM.

Sparsity in DNN models can be exploited by accelerators \cite{han2016eie, parashar2017scnn}. Utilizing sparsity in DNN models for \nc{} is a promising direction for future work.

\section{Conclusion}
Caches have traditionally served only as intermediate low-latency storage units. Our work directly challenges
this conventional design paradigm, and proposes to impose a dual responsibility on caches: store {\em and} compute data. By doing so, we turn them into massively parallel vector units, and drastically reduce on-chip data movement overhead.
In this paper we propose the \nc{} architecture to allow massively parallel compute for Deep Neural Networks. Our advancements in compute cache arithmetic and neural network data layout solutions allow us to provide competitive performance comparably to modern GPUs with negligible area overheads. Nearly three-fourth of a server class processor die area today is devoted for caches. Even accelerators use large caches. Why would one {\em not} want to turn them into compute units?

\section{Acknowledgements}
We thank members of M-Bits research group 
for their feedback. This work was supported in part by the NSF CAREER-1652294 award, and Intel gift award.


\renewcommand{\baselinestretch}{0.84} 
\newcommand{\BIBdecl}{\setlength{\itemsep}{0.25 em}}
\bibliographystyle{IEEEtran}
\bibliography{ref}

\begin{thebibliography}{10}
\providecommand{\url}[1]{#1}
\csname url@samestyle\endcsname
\providecommand{\newblock}{\relax}
\providecommand{\bibinfo}[2]{#2}
\providecommand{\BIBentrySTDinterwordspacing}{\spaceskip=0pt\relax}
\providecommand{\BIBentryALTinterwordstretchfactor}{4}
\providecommand{\BIBentryALTinterwordspacing}{\spaceskip=\fontdimen2\font plus
\BIBentryALTinterwordstretchfactor\fontdimen3\font minus
  \fontdimen4\font\relax}
\providecommand{\BIBforeignlanguage}[2]{{%
\expandafter\ifx\csname l@#1\endcsname\relax
\typeout{** WARNING: IEEEtran.bst: No hyphenation pattern has been}%
\typeout{** loaded for the language `#1'. Using the pattern for}%
\typeout{** the default language instead.}%
\else
\language=\csname l@#1\endcsname
\fi
#2}}
\providecommand{\BIBdecl}{\relax}
\BIBdecl

\bibitem{mwall}
W.~A. Wulf and S.~A. McKee, ``Hitting the memory wall: Implications of the
  obvious,'' \emph{SIGARCH Comput. Archit. News}, vol.~23, no.~1, pp. 20--24,
  Mar. 1995.

\bibitem{Gokhale95}
M.~Gokhale, B.~Holmes, and K.~Iobst, ``Processing in memory: The terasys
  massively parallel pim array,'' \emph{Computer}, vol.~28, no.~4, pp. 23--31,
  1995.

\bibitem{Kogge94}
P.~M. Kogge, ``Execube-a new architecture for scaleable mpps,'' in
  \emph{Parallel Processing, 1994. Vol. 1. ICPP 1994. International Conference
  on}, vol.~1.\hskip 1em plus 0.5em minus 0.4em\relax IEEE, 1994, pp. 77--84.

\bibitem{Patterson97}
D.~Patterson, T.~Anderson, N.~Cardwell, R.~Fromm, K.~Keeton, C.~Kozyrakis,
  R.~Thomas, and K.~Yelick, ``A case for intelligent ram,'' \emph{Micro, IEEE},
  vol.~17, no.~2, pp. 34--44, 1997.

\bibitem{Hmc14}
``Hybrid memory cube specification 2.0,'' 2014.

\bibitem{Ahn15}
J.~Ahn, S.~Yoo, O.~Mutlu, and K.~Choi, ``Pim-enabled instructions: a
  low-overhead, locality-aware processing-in-memory architecture,'' in
  \emph{Computer Architecture (ISCA), 2015 ACM/IEEE 42nd Annual International
  Symposium on}.\hskip 1em plus 0.5em minus 0.4em\relax IEEE, 2015, pp.
  336--348.

\bibitem{neurocube-isca16}
D.~Kim, J.~Kung, S.~Chai, S.~Yalamanchili, and S.~Mukhopadhyay, ``Neurocube: A
  programmable digital neuromorphic architecture with high-density 3d memory,''
  in \emph{Proceedings of ISCA}, vol.~43.\hskip 1em plus 0.5em minus
  0.4em\relax IEEE, 2016, pp. 380--392.

\bibitem{jeloka201628}
S.~Jeloka, N.~B. Akesh, D.~Sylvester, and D.~Blaauw, ``A 28 nm configurable
  memory (tcam/bcam/sram) using push-rule 6t bit cell enabling
  logic-in-memory,'' \emph{IEEE Journal of Solid-State Circuits}, vol.~51,
  no.~4, pp. 1009--1021, 2016.

\bibitem{cc-hpca17}
S.~Aga, S.~Jeloka, A.~Subramaniyan, S.~Narayanasamy, D.~Blaauw, and R.~Das,
  ``Compute caches,'' in \emph{Proceedings of the 23rd International Symposium
  on High Performance Computer Architecture (HPCA-23)}.\hskip 1em plus 0.5em
  minus 0.4em\relax IEEE, 2017, pp. 481--492.

\bibitem{cong2014minimizing}
J.~Cong and B.~Xiao, ``Minimizing computation in convolutional neural
  networks,'' in \emph{International conference on artificial neural
  networks}.\hskip 1em plus 0.5em minus 0.4em\relax Springer, 2014, pp.
  281--290.

\bibitem{chen2016eyeriss}
Y.-H. Chen, J.~Emer, and V.~Sze, ``Eyeriss: A spatial architecture for
  energy-efficient dataflow for convolutional neural networks,'' in
  \emph{Computer Architecture (ISCA), 2016 ACM/IEEE 43rd Annual International
  Symposium on}.\hskip 1em plus 0.5em minus 0.4em\relax IEEE, 2016, pp.
  367--379.

\bibitem{Jeloka15}
S.~Jeloka, N.~Akesh, D.~Sylvester, and D.~Blaauw, ``A configurable tcam / bcam
  / sram using 28nm push-rule 6t bit cell,'' ser. IEEE Symposium on VLSI
  Circuits.\hskip 1em plus 0.5em minus 0.4em\relax IEEE, 2015, pp. C272--C273.

\bibitem{Kang15}
M.~Kang, E.~P. Kim, M.~s.~Keel, and N.~R. Shanbhag, ``Energy-efficient and high
  throughput sparse distributed memory architecture,'' in \emph{2015 IEEE
  International Symposium on Circuits and Systems (ISCAS)}.\hskip 1em plus
  0.5em minus 0.4em\relax IEEE, 2015, pp. 2505--2508.

\bibitem{xeonllc45nm}
M.~Huang, M.~Mehalel, R.~Arvapalli, and S.~He, ``An energy efficient 32-nm
  20-mb shared on-die {L3} cache for intel{\textregistered}
  xeon{\textregistered} processor {E5} family,'' \emph{J. Solid-State
  Circuits}, vol.~48, no.~8, pp. 1954--1962, 2013.

\bibitem{xeonllc22nm}
W.~Chen, S.-L. Chen, S.~Chiu, R.~Ganesan, V.~Lukka, W.~W. Mar, and S.~Rusu, ``A
  22nm 2.5 mb slice on-die l3 cache for the next generation
  xeon{\textregistered} processor,'' in \emph{VLSI Technology (VLSIT), 2013
  Symposium on}.\hskip 1em plus 0.5em minus 0.4em\relax IEEE, 2013, pp.
  C132--C133.

\bibitem{bitserial1}
K.~E. Batcher, ``Bit-serial parallel processing systems,'' \emph{IEEE
  Transactions on Computers}, vol.~31, no.~5, pp. 377--384, 1982.

\bibitem{bitserial2}
P.~B. Denyer and D.~Renshaw, \emph{VLSI signal processing: a bit-serial
  approach}, vol.~1.

\bibitem{ca-micro17}
A.~Subramaniyan, J.~Wang, E.~Balasubramanian, D.~Blaauw, D.~Sylvester, and
  R.~Das, ``Cache automaton,'' in \emph{Proceedings of the 50th Annual IEEE/ACM
  International Symposium on Microarchitecture}.\hskip 1em plus 0.5em minus
  0.4em\relax ACM, 2017, pp. 259--272.

\bibitem{gupta2015deep}
S.~Gupta, A.~Agrawal, K.~Gopalakrishnan, and P.~Narayanan, ``Deep learning with
  limited numerical precision,'' in \emph{Proceedings of the 32nd International
  Conference on Machine Learning (ICML-15)}, 2015, pp. 1737--1746.

\bibitem{han2015deep}
S.~Han, H.~Mao, and W.~J. Dally, ``Deep compression: Compressing deep neural
  networks with pruning, trained quantization and huffman coding,'' \emph{arXiv
  preprint arXiv:1510.00149}, 2015.

\bibitem{zhou2016dorefa}
S.~Zhou, Y.~Wu, Z.~Ni, X.~Zhou, H.~Wen, and Y.~Zou, ``Dorefa-net: Training low
  bitwidth convolutional neural networks with low bitwidth gradients,''
  \emph{arXiv preprint arXiv:1606.06160}, 2016.

\bibitem{jouppi2017datacenter}
N.~P. Jouppi, C.~Young, N.~Patil, D.~Patterson \emph{et~al.}, ``In-datacenter
  performance analysis of a tensor processing unit,'' in \emph{Proceedings of
  the 44th Annual International Symposium on Computer Architecture}.\hskip 1em
  plus 0.5em minus 0.4em\relax ACM, 2017, pp. 1--12.

\bibitem{parabixtrans}
``{Parabix Transform.}''
  \url{http://parabix.costar.sfu.ca/wiki/ParabixTransform}, accessed:
  2017-11-20.

\bibitem{parabix}
D.~Lin, N.~Medforth, K.~S. Herdy, A.~Shriraman, and R.~Cameron, ``Parabix:
  Boosting the efficiency of text processing on commodity processors,'' in
  \emph{High Performance Computer Architecture (HPCA), 2012 IEEE 18th
  International Symposium on}.\hskip 1em plus 0.5em minus 0.4em\relax IEEE,
  2012, pp. 1--12.

\bibitem{intelcat}
{Intel Corporation}, ``{Cache Allocation Technology},''
  \url{https://software.intel.com/en-us/articles/introduction-to-cache-allocation-technology},
  accessed: 2017-11-20.

\bibitem{szegedy2016rethinking}
C.~Szegedy, V.~Vanhoucke, S.~Ioffe, J.~Shlens, and Z.~Wojna, ``Rethinking the
  inception architecture for computer vision,'' in \emph{Proceedings of the
  IEEE Conference on Computer Vision and Pattern Recognition}, 2016, pp.
  2818--2826.

\bibitem{david2010rapl}
H.~David, E.~Gorbatov, U.~R. Hanebutte, R.~Khanna, and C.~Le, ``Rapl: memory
  power estimation and capping,'' in \emph{Low-Power Electronics and Design
  (ISLPED), 2010 ACM/IEEE International Symposium on}.\hskip 1em plus 0.5em
  minus 0.4em\relax IEEE, 2010, pp. 189--194.

\bibitem{nvidiasmi}
{Nvidia Corporation}, ``Nvidia system management interface,''
  \url{https://developer.nvidia.com/nvidia-system-management-interface},
  accessed: 2017-11-18.

\bibitem{intelvtune}
{Intel Corporation}, ``Intel vtune amplifier performance profiler,''
  \url{https://software.intel.com/en-us/intel-vtune-amplifier-xe}, accessed:
  2017-11-18.

\bibitem{seshadri2017ambit}
V.~Seshadri, D.~Lee, T.~Mullins, H.~Hassan, A.~Boroumand, J.~Kim, M.~A. Kozuch,
  O.~Mutlu, P.~B. Gibbons, and T.~C. Mowry, ``Ambit: In-memory accelerator for
  bulk bitwise operations using commodity dram technology,'' in
  \emph{Proceedings of the 50th Annual IEEE/ACM International Symposium on
  Microarchitecture}.\hskip 1em plus 0.5em minus 0.4em\relax ACM, 2017, pp.
  273--287.

\bibitem{li2017drisa}
S.~Li, D.~Niu, K.~T. Malladi, H.~Zheng, B.~Brennan, and Y.~Xie, ``Drisa: A
  dram-based reconfigurable in-situ accelerator,'' in \emph{Proceedings of the
  50th Annual IEEE/ACM International Symposium on Microarchitecture}.\hskip 1em
  plus 0.5em minus 0.4em\relax ACM, 2017, pp. 288--301.

\bibitem{gen6-datasheet}
{Intel Corporation}, \emph{6th Generation
  Intel\textsuperscript{\textregistered} Processor Datasheet for S-Platforms},
  2015.

\bibitem{intel-me}
I.~Skochinsky, ``Secrets of intel management engine,''
  \url{http://www.slideshare.net/codeblue_jp/igor-skochinsky-enpub}, accessed:
  2016-02-17.

\bibitem{sgx-me}
S.~Gueron, ``A memory encryption engine suitable for general purpose
  processors.'' \emph{IACR Cryptology ePrint Archive}, vol. 2016, p. 204, 2016.

\bibitem{shafiee2016isaac}
A.~Shafiee, A.~Nag, N.~Muralimanohar, R.~Balasubramonian, J.~P. Strachan,
  M.~Hu, R.~S. Williams, and V.~Srikumar, ``Isaac: A convolutional neural
  network accelerator with in-situ analog arithmetic in crossbars,'' in
  \emph{Proceedings of the 43rd International Symposium on Computer
  Architecture}.\hskip 1em plus 0.5em minus 0.4em\relax IEEE Press, 2016, pp.
  14--26.

\bibitem{chi2016prime}
P.~Chi, S.~Li, C.~Xu, T.~Zhang, J.~Zhao, Y.~Liu, Y.~Wang, and Y.~Xie, ``Prime:
  a novel processing-in-memory architecture for neural network computation in
  reram-based main memory,'' in \emph{ACM SIGARCH Computer Architecture News},
  vol.~44, no.~3.\hskip 1em plus 0.5em minus 0.4em\relax IEEE Press, 2016, pp.
  27--39.

\bibitem{chen2014dadiannao}
Y.~Chen, T.~Luo, S.~Liu, S.~Zhang, L.~He, J.~Wang, L.~Li, T.~Chen, Z.~Xu,
  N.~Sun, and O.~Temam, ``Dadiannao: A machine-learning supercomputer,'' in
  \emph{Proceedings of the 47th Annual IEEE/ACM International Symposium on
  Microarchitecture}.\hskip 1em plus 0.5em minus 0.4em\relax IEEE Computer
  Society, 2014, pp. 609--622.

\bibitem{chung2017accelerating}
E.~Chung, J.~Fowers, K.~Ovtcharov, M.~Papamichael, A.~Caulfield, T.~Massengil,
  M.~Liu, D.~Lo, S.~Alkalay, M.~Haselman, C.~Boehn, A.~Forin, K.~S. Gatlin,
  M.~Ghandi, S.~Heil, K.~Holohan, T.~Juhasz, R.~K. Kovvuri, S.~Lanka, F.~v.
  Megen, D.~Mukhortov, P.~Patel, S.~Reinhardt, A.~Sapek, R.~Seera,
  B.~Sridharan, L.~Woods, P.~Yi-Xiao, R.~Zhao, and D.~Burger, ``Accelerating
  persistent neural networks at datacenter scale,'' 2017, hot Chips: A
  Symposium on High Performance Chips.

\bibitem{gokhale1995processing}
M.~Gokhale, B.~Holmes, and K.~Iobst, ``Processing in memory: The terasys
  massively parallel pim array,'' \emph{Computer}, vol.~28, no.~4, pp. 23--31,
  1995.

\bibitem{judd2016stripes}
P.~Judd, J.~Albericio, T.~Hetherington, T.~M. Aamodt, and A.~Moshovos,
  ``Stripes: Bit-serial deep neural network computing,'' in
  \emph{Microarchitecture (MICRO), 2016 49th Annual IEEE/ACM International
  Symposium on}.\hskip 1em plus 0.5em minus 0.4em\relax IEEE, 2016, pp. 1--12.

\bibitem{han2016eie}
S.~Han, X.~Liu, H.~Mao, J.~Pu, A.~Pedram, M.~A. Horowitz, and W.~J. Dally,
  ``Eie: efficient inference engine on compressed deep neural network,'' in
  \emph{Proceedings of the 43rd International Symposium on Computer
  Architecture}.\hskip 1em plus 0.5em minus 0.4em\relax IEEE Press, 2016, pp.
  243--254.

\bibitem{parashar2017scnn}
A.~Parashar, M.~Rhu, A.~Mukkara, A.~Puglielli, R.~Venkatesan, B.~Khailany,
  J.~Emer, S.~W. Keckler, and W.~J. Dally, ``Scnn: An accelerator for
  compressed-sparse convolutional neural networks,'' in \emph{Proceedings of
  the 44th Annual International Symposium on Computer Architecture}.\hskip 1em
  plus 0.5em minus 0.4em\relax ACM, 2017, pp. 27--40.

\end{thebibliography}

\end{document}